\documentclass[12pt,a4paper]{article}
\usepackage[OT2,T1]{fontenc}
\usepackage{amssymb,amsmath,stmaryrd,mathtools,bm,graphicx,caption,subcaption,here,parskip,tikz-cd,tikz,cancel,ulem,cite,titlesec,enumerate,empheq,changepage,physics,cancel}
\usepackage{tikz-3dplot}

\usepackage[rightcaption]{sidecap}
\usepackage[margin=.9in]{geometry}
\usepackage[most]{tcolorbox}
\usepackage[colorlinks,allcolors=blue]{hyperref}
\DeclareMathAlphabet{\mathscrbf}{OMS}{mdugm}{b}{n}
\usetikzlibrary{decorations.markings,decorations.pathreplacing}
\newtcbox{\othermathbox}[1][]{nobeforeafter, math upper, tcbox raise base, 
          enhanced, rounded corners, colback=black!5, colframe=black,
          left=0.7em, top=0.4em, right=0.7em, bottom=0.5em}

\definecolor{MyYellow}{RGB}{248,199,82}
\definecolor{mygreen}{rgb}{0.5,0.71,0.5}
\definecolor{myorange}{rgb}{1,0.5,0}
\makeatletter
\newcommand{\ostar}{\mathbin{\mathpalette\make@circled\star}}
\newcommand{\make@circled}[2]{%
  \ooalign{$\m@th#1\smallbigcirc{#1}$\cr\hidewidth$\m@th#1#2$\hidewidth\cr}%
}
\newcommand{\smallbigcirc}[1]{%
  \vcenter{\hbox{\scalebox{0.77778}{$\m@th#1\bigcirc$}}}%
}
\makeatother

\renewcommand{\i}[1]{\textit{#1}}
\newcommand{\cF}{F}
\numberwithin{equation}{section}

\begin{document}

\begingroup
\renewcommand{\thefootnote}{\fnsymbol{footnote}}
\setcounter{footnote}{0}
\begin{center}
{\LARGE{{Quantum Mechanics on Lie Groups:\\[.4em]
I.~Noncommutative Fourier Transforms}}}\\[1.3em]

{\large Mathieu Beauvillain,$^1$\footnotemark[1] Blagoje Oblak,$^2$ and Marios Petropoulos$^1$}\\[1.3em]

{\small%
$^1$ CPHT, CNRS, École polytechnique, Institut Polytechnique de Paris, 91120 Palaiseau, France\\
$^2$ Université Claude Bernard Lyon 1, ICJ UMR 5208, CNRS, 69622 Villeurbanne, France}\\
\end{center}
~\\[2em]

\begin{center}
\begin{minipage}{.8\textwidth}
\textbf{Abstract.} Starting from square-integrable wave functions on a Lie group, we build an invertible Fourier transform mapping them on wave functions on the dual of the Lie algebra. This is a group-theoretic version of the map from position space to momentum space, with generally noncommuting momenta owing to the group structure. As a result, the multiplication of momentum-dependent functions involves star products, which makes the construction of noncommutative Fourier series much more involved than that of their commutative cousin. This is especially true when compact subgroups are present, in which case we carefully take into account quotients of the operator algebra, and the resulting normalization issues. We show that our formalism provides an isometry of Hilbert spaces, and use it to derive a noncommutative Poisson summation formula for any compact Lie group. This is a key preliminary for the computation of Wigner functions and path integrals for quantum systems on group manifolds.
\end{minipage}
\end{center}
\footnotetext[1]{Corresponding author. Contact at mathieu.beauvillain@polytechnique.edu}
\endgroup

\newpage
\tableofcontents

\newpage
\section{Motivation and summary}
\label{sec: intro}

Many physical systems have configuration spaces given by a Lie group---possibly an infinite-dimensional one. Prime examples are rigid bodies \cite{Montgomery}, spin chains \cite{LandauLifschitz} and fluid flows \cite{ArnoldOrigin}, respectively corresponding to rotation groups, loop groups and diffeomorphism groups. In all those cases, the group structure provides powerful geometric tools for physical predictions and/or numerical simulations (see \i{e.g.}~\cite{cotter2018numerically}), typically thanks to the fact that solutions of the equations of motion are geodesics in the group. Such dynamics is described by Euler-Arnold equations or, more generally, Lie-Poisson equations \cite{ArnoldKhesin,Khesin}.

Surprisingly, much remains to be understood about the \i{quantization} of Lie-Poisson systems. This is so despite many instances of quantum problems that explicitly rely on a group structure. Referring again to the examples above, quantum rotors are essential for the energy spectrum of molecules \cite{Allen,Gripaios:2015pfa}, spin chains form the foundation of quantum integrability \cite{gaudin2014}, and quantum liquids play a key role in statistical and condensed matter physics \cite{Leggett,Delacretaz}. It is therefore desirable to develop a general approach to quantum mechanics on Lie groups, akin to what is already well known for simpler systems on $\mathbb{R}^n$.

The goal of the present paper and its follow-up \cite{Beauvillain2} is to establish this framework. Broadly speaking, the plan is to consider Hilbert spaces of the form $L^2(G)$, where $G$ is a Lie group, and build the standard tools of quantum mechanics in that context. This crucially includes Wigner functions \cite{Zachos,StPierre,Mukunda:2003wna,Mukunda:2005us}, path integrals \cite{Jevicki:1978yv,Krausz:1997gw}, and the reduction to coadjoint orbits relevant for generalized coherent states \cite{Perelomov:1986uhd,Figueroa:1990zu}. However, a preliminary requirement for all those considerations is to understand the passage from `position space' to `momentum space' through a Fourier transform. This is the specific issue addressed here. Remarkably, the same question was raised two decades ago in the context of (loop) quantum gravity \cite{Freidel:2005bb,Joung:2008mr,Oriti:2011ac,Raasakka:2011np,Guedes:2013vi}, whose configuration space consists of a thermodynamically large product of rotation groups. Our approach is ultimately similar to the one developed there, but differs in key details, mostly having to do with the proper treatment of quotients that stem from compact directions in $G$.

\paragraph{Momentum space dual to a Lie group.} There are two obvious answers to what `momentum space' means when `position space' is a Lie group $G$. Provided the latter is compact, the Peter-Weyl theorem \cite{Williams} (see also \cite[sec.~3]{Sepanski}) states that the regular representation of $G$ in $L^2(G)$ decomposes into a direct sum of irreducible representations of $G$, each weighed by its dimension. This effectively provides Fourier series for functions on $G$, with orthonormal harmonics given by the matrix elements of irreducible representations. It is thus tempting to call `momentum space' the (discrete) set of such matrix elements and irreducible representations, in the same way that Fourier modes on a circle are labelled by an integer. Equivalently, one may choose to call `momentum space' the union of integral coadjoint orbits of $G$ \cite{Figueroa:1990zu,Wildberger}, as is sometimes done in noncommutative geometry \cite{Hammou:2001cc}. The advantage of this approach is that Fourier series are explicitly given by the Peter-Weyl theorem; its drawback is that it appears naturally neither for Wigner functions nor for path integrals. Indeed, the latter call instead for fully continuous coordinates in the classical phase space $T^*G$, \i{i.e.}~the cotangent bundle of $G$. This is equivalent to the product $G\times\mathfrak{g}^*$, where $\mathfrak{g}^*$ is the entire dual space of the Lie algebra $\mathfrak{g}$ of $G$. Thus, a second option is to call `momentum space' the vector space $\mathfrak{g}^*$, as is indeed routinely done when dealing with Lie-Poisson equations \cite{ArnoldKhesin,Khesin}. The advantage then is that the link with classical physics is manifest; the drawback is that Fourier transforms are much more involved.

There is nothing wrong with either choice; each has advantages and drawbacks that depend on context. For the case of Wigner functions, path integrals and their (semi)classical limits, the most appealing approach is the second one: to call `momentum space' the dual space $\mathfrak{g}^*$ of $\mathfrak{g}$. The question is how exactly one is supposed to take Fourier transforms to a noncommutative momentum space \cite{Oriti:2011ac,Raasakka:2011np,Guedes:2013vi} (see also \cite{kapranov}), and how this relates to the Peter-Weyl decomposition, more natural from a representation-theoretic perspective. (Wigner functions on curved configuration spaces, but with \i{commuting} momenta, were considered in \cite{Fischer:2013pch}; we do not follow this approach.)

Noncommutative Fourier transforms were initially introduced in loop quantum gravity \cite{Freidel:2005bb,Oriti:2011ac,Raasakka:2011np,Guedes:2013vi}, precisely with the goal of deriving semiclassical dynamics in the fully continuous phase space $T^*G$ of a quantum system whose configuration space is a (very large) Lie group $G$. Here, we similarly define noncommutative Fourier transforms to a fully continuous momentum space, but our construction differs in several ways from those in \cite{Oriti:2011ac,Raasakka:2011np,Guedes:2013vi}. We emphasize this at several points in the text below, but perhaps the most notable distinction is our careful treatment of the quotients that arise in momentum space $\mathfrak{g}^*$ when $G$ has compact subgroups, and of the way in which this interferes with noncommutativity. For instance, the procedure in \cite{Oriti:2011ac,Raasakka:2011np,Guedes:2013vi} defined the space of wave functions in momentum space as the image of $L^2(G)$ by the Fourier transform. We will instead use a different definition of momentum-space wave functions, irrespective of the Fourier transform, then show that Fourier coefficients and their inverse, Fourier series, actually provide isometries of Hilbert spaces. Relatedly, our approach avoids the use of equivalence classes of functions, which turns out to make the whole construction well-defined.\footnote{With the conventions of \cite{Oriti:2011ac,Raasakka:2011np,Guedes:2013vi}, the action of operators on the momentum Hilbert space explicitly depends on one's choice of representative in an equivalence class, and is therefore ill-defined. Restoring well-definiteness for the action of operators naturally led us to the similar, but ultimately different, construction outlined in this work.} All such seemingly technical details are crucial in practice for concrete, analytical computations in the semiclassical regime, which will be further explored in \cite{Beauvillain2}.

\paragraph{Plan.} The paper is organized as follows. In section \ref{sec: operator algebra}, we briefly review the symplectic structure of $T^*G$ and use it to build the operator algebra relevant for quantized Lie-Poisson systems. Section \ref{sec: representations} is devoted to `position space' and `momentum space' representations of this algebra, with an emphasis on the quotients needed when $G$ has compact subgroups, and on their delicate representation-theoretic consequences. Section \ref{sec:DefFourier} links position and momentum representations through a unitary noncommutative Fourier transform, which we define and investigate in detail. In particular, we carefully discuss the distinction between Fourier \i{transforms} and Fourier \i{coefficients} or \i{series}. This is crucial for any concrete computation that relies on noncommutative Fourier transforms for an exponential Lie group, but to our knowledge it has never been discussed in the literature. In section \ref{sec: examples}, we apply the tools of noncommutative Fourier series to two simple examples: the Abelian group $G=\text{U(1)}$ and the nonabelian $G=\text{SU(2)}$. For the latter, we show that our framework reproduces the Kirillov character formula \cite{KirillovLectures}. We also derive a nonabelian Poisson summation formula, which we generalize to any Lie group. Finally, the short section \ref{Duflo map} revisits the construction of noncommutative Fourier series for the so-called Duflo quantization prescription of momentum operators \cite{Duflo}. We show there, in particular, that the Parseval-Plancherel formula for class functions follows from noncommmutative Fourier series, and that the character of an irreducible representation has a single Fourier coefficient, localized on a coadjoint orbit of the group.

A word of caution may be appropriate at this point. While the content of this paper is mathematical, it is not phrased in the rigorous language dear to mathematicians. The presentation is, instead, geared towards physicists, to whom it is chiefly addressed. In particular, we assume throughout that the group $G$ is finite-dimensional and weakly exponential, meaning that the exponential from the Lie algebra to the Lie group is at least dense. We will also encounter infinite factors and singular distributions, which are ubiquitous (and expected) when dealing with plane waves in quantum mechanics. Furthermore, we systematically focus on smooth functions and wave functions (unless explicitly stated otherwise). We expect our results to admit suitable generalizations to infinite-dimensional Lie-Fr\'echet groups such as those alluded to above (see \textit{e.g.}~\cite{Khesin, ArnoldKhesin}), but no attempt is made here to rigorously obtain these generalizations.

\section{Operator algebras on Lie groups}
\label{sec: operator algebra}

In this section, we construct the operator algebra of quantum mechanics on Lie groups through canonical quantization. We first review the Poisson bracket of functions on the phase space $T^*G$, then build an operator algebra by `putting hats' on functions, and replacing Poisson brackets by commutators. The key subtlety resides in the fact that momentum operators fail to commute. This calls for an ordering prescription, which we choose to be symmetric to ensure later compatibility of the star product with exponentials.

\subsection{Poisson brackets}

Consider a mechanical system whose configuration space is a Lie group $G$, with elements $g$, $h$, etc. Call $\mathfrak{g}$ the Lie algebra of $G$, viewed as the space of right-invariant vector fields on $G$;\footnote{Our convention to define the Lie algebra as consisting of \textit{right}-invariant vector fields is motivated by the fact that they generate infinitesimal \textit{left} translations when acting on functions.} denote Lie algebra elements as $X$, $Y$, etc. Let $\exp:\mathfrak{g}\rightarrow G$ be the exponential map, understood as the time-1 flow of such vector fields starting from the identity. For later use, choose a basis $\{t_i|i=1,...,n\}$ of $\mathfrak{g}$, so that each $t_i$ gives rise to a Lie derivative operator ${\cal L}_i:={\cal L}_{t_i}$ acting on functions on $G$. The Lie bracket of generators reads $[t_i,t_j]=c_{ij}{^k}t_k$, with structure constants $c_{{i}j}{}^k$ and implicit summation over repeated indices.

\paragraph{Phase space.} The corresponding phase space is the cotangent bundle $T^*G$, which turns out to be trivial: it is globally equivalent to the direct product $G\times\mathfrak{g}^*$, where $\mathfrak{g}^*$ is the dual vector space of the Lie algebra $\mathfrak{g}$. (See \i{e.g.}~\cite[Appendix A]{OblakKozy} for the proof of this equivalence.) We denote `momenta' as $p,q\in\mathfrak{g}^*$ and write each of them as $p=p_i(t^i)^*$ in terms of the dual basis such that $\langle(t^i)^*,t_j\rangle=\delta^i_j$.\footnote{The pairing $\langle\cdot,\cdot\rangle$ is not a scalar product: it is just the pairing between a vector space and its dual, here $\mathfrak{g}$ and $\mathfrak{g}^*$. Scalar products of wave functions will instead be denoted $\langle\cdot|\cdot\rangle$.} The phase space is endowed with a symplectic form whose inverse defines the Poisson bracket of functions on $G\times\mathfrak{g}^*$, namely
\begin{align}
      \{A,B\}
      =
      \mathcal{L}_i A\frac{\partial B}{\partial p_i} -
      \frac{\partial A}{\partial p_i}\mathcal{L}_i B - c_{ij}{}^k \frac{\partial A}{\partial p_i}\frac{\partial B}{\partial p_j}p_k
      \label{eq: Lie Poisson bracket}
\end{align}
for any two functions $A(g,p)$ and $B(g,p)$. In particular, this applies to functions locally given by coordinates $g^i$ on $G$ and momentum components $p_i$ on $\mathfrak{g}^*$, for which\footnote{Coordinate functions are typically not globally smooth on $G$. The brackets \eqref{eq: component PB G} involving $g^i$s thus hold on open neighbourhoods where the $g^i$s are smooth.}
\begin{align}
\{g^i,g^j\} = 0,
\qquad \{g^j , p_i\} = \mathcal{L}_i g^j,
\qquad \{p_i,p_j\} = -c_{ij}{}^k p_k.
\label{eq: component PB G}
\end{align}
We denote by $C^{\infty}(G\times\mathfrak{g}^*)$ the Poisson algebra of (smooth, complex-valued) functions on $G\times\mathfrak{g}^*$ endowed with the Poisson bracket \eqref{eq: Lie Poisson bracket}. The key difference between this algebra and the more standard one of functions on $T^*\mathbb{R}^n$ is that momenta fail to commute in \eqref{eq: component PB G}. Put differently, momentum space is generally not a Lagrangian submanifold of phase space, so it admits no geometric quantization \cite{woodhouse1997}. This is the main difficulty in dealing with quantum mechanics on Lie groups.

\begin{figure}
    \centering
    \includegraphics[width=.4\textwidth]{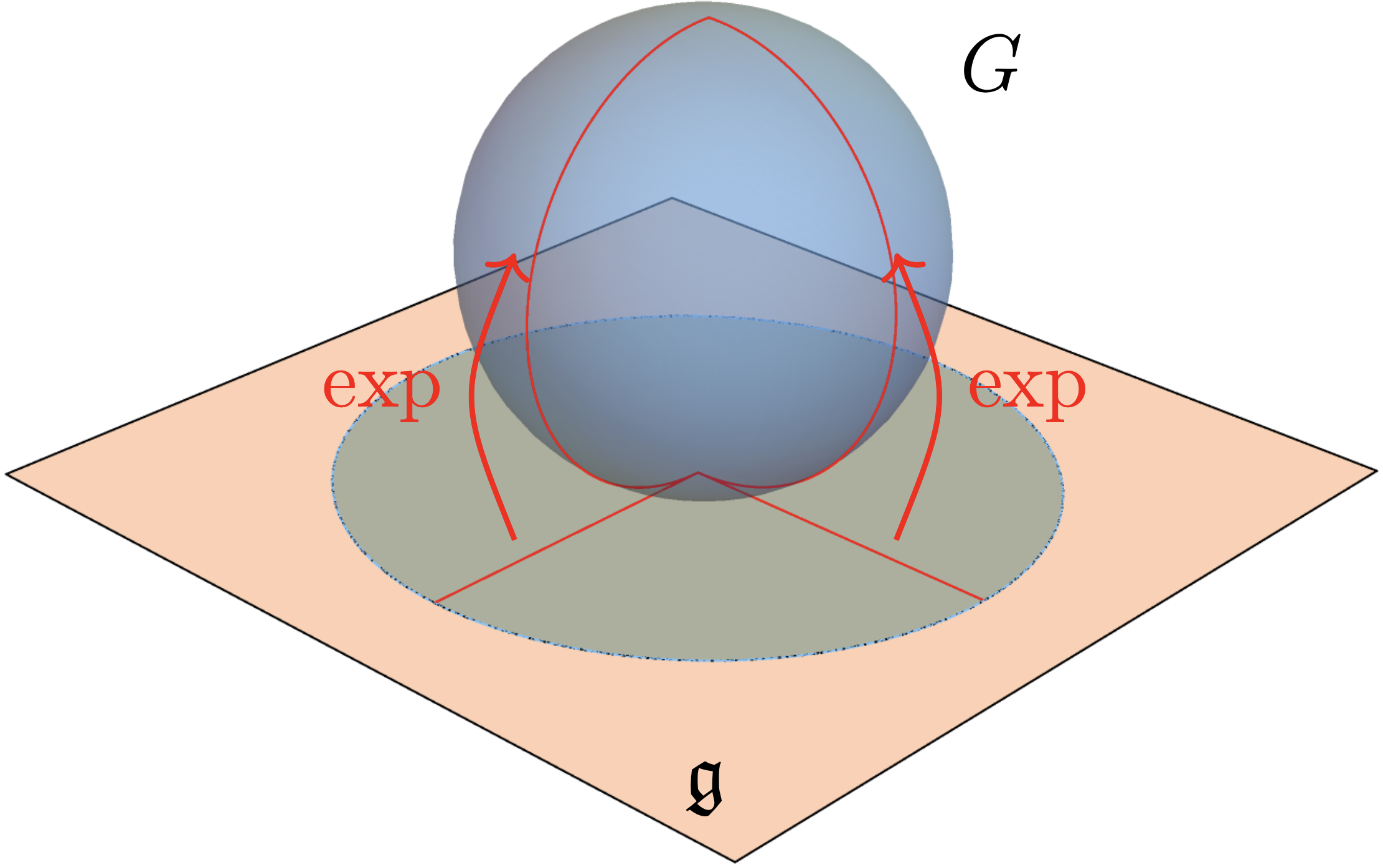}
\caption{A cartoon of the principal branch (blue disk) of the logarithm from a Lie algebra $\mathfrak{g}$ (the plane) to a Lie group $G$ (the sphere). On the principal branch, the exponential is injective, and its image in the group is dense. In the picture, the exponential of the principal branch only misses the north pole. The exponential is not injective on the boundary of the principal branch; this boundary is a circle in the present case, wholly sent on the north pole by the exponential. In fact, this exact picture is stricly valid for $G=\text{SU(2)}$, up to the replacement of $S^2$ by $S^3$ and $\mathbb{R}^2$ by $\mathbb{R}^3$: see section \ref{sec: examples}.}
    \label{fig:principalbranch}
\end{figure}

\paragraph{Principal branch coordinates on $\boldsymbol G$.} Our aim is to quantize the Poisson algebra of functions on $G\times\mathfrak{g}^*$. In doing so, we will face the problem of defining `position operators', which is complicated by the fact that $G$ need not admit global coordinates. We therefore rely on the Lie algebra and the exponential map to define coordinates on the group. Namely, define the \i{principal branch} of the logarithm to be the open set in $\mathfrak{g}$ that contains $0$, for which the exponential is both injective and dense in $G$, and that is `symmetric' so that if $X$ belongs to the principal branch, then $-X$ does as well; see fig.~\ref{fig:principalbranch}. Coordinates $X^i(g)$ on $G$ are then chosen such that $g=\exp(X^i(g)t_i)$ in terms of the basis $\{t_i|i=1,...,n\}$ introduced above. By construction, these coordinates are dense in $G$, so they uniquely label almost any $g\in G$. They are such that the identity $e\in G$ sits at the origin $X^i(e)=0$, and such that the inverse of an element in $G$ is the opposite vector in coordinates: $X^i(g^{-1})=-X^i(g)$. Furthermore, they satisfy
\begin{align}
       \mathcal{L}_i X^j
       =
       \left.\frac{\partial}{\partial {Y^i}}\right|_{Y=0}B\big(Y^kt_k,X^{\ell}t_{\ell}\big)^j,
       \label{eq: prop of pb coordinates}
\end{align}
where $B(Y,X)$ denotes the Baker-Campbell-Hausdorff expansion. The latter is defined so that, for any two Lie algebra elements $X,Y$, one has $e^{B(X,Y)}:=e^Xe^Y$ in the (closure of the) universal enveloping algebra of $\mathfrak{g}$, which explicitly yields the usual series \cite{Hall}
\begin{align}
B(Y,X)
=
Y+X+\frac{1}{2}[Y,X]+\frac{1}{12}\left([Y,[Y,X]]-[X,[X,Y]]\right)+\cdots.
\label{baker}
\end{align}
As a special case of eq.~\eqref{eq: prop of pb coordinates} note that $\mathcal{L}_i X^j(e)=\delta_i^j$ at the identity, which will be useful below. We stress that \eqref{baker} does not rely on any exponential map from the Lie algebra $\mathfrak{g}$ to the group $G$: the definition of $B(X,Y)$ is purely algebraic and solely relies on commutation relations of $\mathfrak{g}$ used in the universal enveloping algebra.\footnote{%
The Baker-Campbell-Hausdorff expansion \eqref{baker} can notoriously fail to converge, in which case $B(X,Y)$ is defined by `lifting the group law to the Lie algebra'. Specifically, let $\gamma(t)$ be the path in $\mathfrak{g}$ such that $\exp(\gamma(t)) = \exp(tX)\exp(tY)$, and such that it intersects the loci where the exponential fails to be injective transversely. Then define $B(X,Y) := \gamma(1)$.
} 

\paragraph{Time evolution.} We will not be concerned with quantum dynamics for now \cite{Krausz:1997gw}, but it is worth briefly recalling its classical version, to which we return at the end of section \ref{sec:DefFourier}. As in any Hamiltonian system, the time evolution of a function on $G\times\mathfrak{g}^*$ is given by its Poisson bracket with a Hamiltonian function $H(g,p)$. This leads to a noncommutative cousin of the standard Hamilton equations of classical mechanics on $T^*\mathbb{R}^n$: given a path $(g_t,p_t)$ in phase space, it solves the equations of motion if
\begin{align}
\label{3a}
      \partial_t gg ^{-1}\,  &= \partial_p H,\\
      \partial_t p - \mathrm{ad}^*_{\partial_t g g ^{-1}}p &= - R_g^* \partial_g H.
      \label{3b}
\end{align}
Let us unpack each term. On the left-hand side of \eqref{3a}, $\partial_t gg ^{-1}$ is a Lie algebra element given by the (right) Maurer-Cartan form on $G$ acting on the tangent vector $\partial_tg$ at $g$. On the right-hand side of \eqref{3a}, $\partial_pH$ is the exterior derivative of $H(g,p)$ in momentum space, \i{i.e.}~the differential of the function $H(g,\cdot)$ on $\mathfrak{g}^*$. It is also a Lie algebra element, since $\partial_pH$ is a linear form on $\mathfrak{g}^*$, \i{i.e.}~an element of $(\mathfrak{g}^*)^*=\mathfrak{g}$. On the left-hand side of \eqref{3b}, $\mathrm{ad}^*$ denotes the coadjoint representation of $\mathfrak{g}$, defined by $\langle\mathrm{ad}^*_X(p),Y\rangle:=-\langle p,[X,Y]\rangle$ for any $X,Y\in\mathfrak{g}$ and any $p\in\mathfrak{g}^*$. Finally, on the right-hand side of \eqref{3b}, $\partial_gH$ is the exterior derivative of $H(g,p)$ in position space, \i{i.e.}~the differential of the function $H(\cdot,p)$ on $G$. What appears in \eqref{3b} is the pullback of this differential to the identity by right multiplication $R_g:G\to G:h\mapsto hg$.

In the special case where $H(g,p)=H(p)$ only depends on momenta, we refer to eqs.~\eqref{3a}--\eqref{3b} as \i{Lie-Poisson equations}. These reduce to Euler-Arnold equations in the even more restricted case where $H(p)$ is a quadratic function, whereupon the solution $g_t$ of \eqref{3a} is a geodesic in $G$ with respect to a left-invariant metric. As mentioned in section \ref{sec: intro}, such dynamical systems are ubiquitous in physics; see \i{e.g.}~\cite{ArnoldKhesin,Khesin} for numerous examples. The goal of the present work is to set the stage for the quantization of Lie-Poisson systems, with a view towards quantum liquids \cite{Delacretaz}.

\subsection{From Poisson brackets to operator algebras}
\label{ssec: operator algebra}

We now define the abstract operator algebra meant to describe the canonical quantization of the Poisson algebra of functions on $G\times\mathfrak{g}^*$. Representations of the operator algebra on actual Hilbert spaces will be provided in section \ref{sec: representations}.

\paragraph{Operator algebra and quantization prescription.}  We wish to find operators mimicking the `position' and `momentum' operators of quantum mechanics. The latter are readily obtained by quantizing the momentum components $p_i$ introduced above \eqref{eq: component PB G}, which thus become operators $\hat p_i$. For position operators, we use the principal branch coordinates defined above eq.~\eqref{eq: prop of pb coordinates}. Any function on $G$ can thus be seen as a function on the principal branch, as $f(g)=f(X^i(g))$. We then promote the coordinate functions $X^i(g)$ to operators $\hat X^i$ and define abstract operators $\hat f = f(\hat X^i)$. Owing to the Poisson brackets \eqref{eq: component PB G}, this quantization must be such that the following commutators hold:\footnote{Here and below, we use units such that $\hbar=1$.}
\begin{equation}
\label{5a}
     [\hat X^i,\hat X^j]=0,\qquad
     [\hat X^j,\hat p_i]=i \widehat{\mathcal{L}_i X^j},\qquad
     [\hat p_i,\hat p_j]= -i  c_{ij}{}^k \hat p_k.
\end{equation}
The main complication compared with quantum mechanics on $\mathbb{R}^n$ appears once again in the noncommutative momenta. We denote by ${\cal A}$ the abstract algebra spanned by the $\hat p_i$s and the $\hat X^j$s with commutators \eqref{5a}. Let also ${\cal A}_{\text{pos}}$ and ${\cal A}_{\text{mom}}$ be the subalgebras of $\cal A$ that respectively correspond to $C^{\infty}(G)$ and $C^{\infty}(\mathfrak{g}^*)$. The subalgebra ${\cal A}_{\text{mom}}$, in particular, is (isomorphic to the closure of) the universal enveloping algebra of $\mathfrak{g}$ owing to the last commutator in eqs.~\eqref{5a}.

To be more precise, the procedure of `putting hats on functions' means that we assume the existence of a linear quantization map
\begin{align}
     \mathcal{Q}: C^{\infty}(G\times\mathfrak{g}^*)\rightarrow {\cal A}, A\mapsto\hat A:=\mathcal{Q}(A)
     \label{eq: quantization map}
\end{align}
such that $\mathcal{Q}(\bar A)=(\mathcal{Q}(A))^{\dagger}$, with $\hat X^i:={\cal Q}(X^i)$ and $\hat p_i:={\cal Q}(p_i)$. This map prescribes an ordering for operators. For example, the classical function $p_2p_1 = p_1p_2$ can be quantized to $\hat p_1\hat p_2$, $\hat p_2\hat p_1$ or $\tfrac12(\hat p_1\hat p_2 +\hat p_2\hat p_1)$, each of which is valid, yet different from the others. What is unusual here, compared with $\mathbb{R}^n$, is that the ordering of momenta matters. We specifically choose this ordering to be symmetric among the $p_i$s:
\begin{align}
     \mathcal{Q}(p_{i_1}\dots p_{i_n})
     :=
     \frac{1}{n!}\sum_{\sigma\in\text{Sym}(n)} 
     \hat p_{i_{\sigma(1)}}\dots \hat p_{i_{\sigma(n)}}.
     \label{eq: symmetric ordering}
\end{align}
The key virtue of this prescription is to simplify computations that involve exponentials of momentum operators, which are crucial for Fourier transforms. Indeed, given any Lie algebra element $X\in\mathfrak{g}$, the quantization \eqref{eq: symmetric ordering} maps the exponential function $p\mapsto e^{-i\langle p,X\rangle}$ on the exponential operator
\begin{align}
    \mathcal Q(e^{-i\langle \cdot, X\rangle})
    =
    e^{-i\langle \hat p, X\rangle},
    \label{expo}
\end{align}
where the expansion of the exponential on the right-hand side contains all possible permutations of the $\hat p_i$s. More generally, it turns out that the Fourier transform can be built with any ordering choice $\mathcal Q$ such that $\mathcal{Q}(e^{-i\langle \cdot , X\rangle}) = f(X)e^{-i\langle \hat p, X\rangle}$ for some function $f$. An example that is not symmetric is the Duflo map (see section \ref{Duflo map}). In principle, infinitely many orderings may be considered, but the symmetric one and the Duflo map are the two known ones that always satisfy the exponential property \eqref{expo} for any Lie group; we focus on the symmetric one throughout, except in section \ref{Duflo map}.

Of course, the full quantization map \eqref{eq: quantization map} also relies on a choice of ordering for products of $X^i$s and $p_i$s, as in the standard Euclidean case. We assume that a choice of ordering has been made, but its details do not matter: they give rise to the same ordering issues as in standard quantum mechanics on $\mathbb{R}^n$ (with phase space $\mathbb{R}^{2n}$).

\paragraph{Exponential operators and branches of the logarithm.} Exponential operators should provide a (possibly projective) representation of $G$ upon identifying $g=\exp(X)$. (Think \i{e.g.}~of representations of SU(2), where exponential operators act as unitary rotations on any Hilbert space.) This is justified by the last commutator in eqs.~\eqref{5a}, which implies that the product of two exponential operators reads
\begin{align}
       e^{-i\langle\hat p, X\rangle}e^{-i\langle \hat p, Y\rangle}
       =
       e^{-i\langle \hat p, B(X,Y)\rangle},
       \label{eq: BCH}
\end{align}
where $B(X,Y)$ is given by the Baker-Campbell-Hausdorff expansion \eqref{baker}. We will denote by $\cal E$ the group of exponential operators \eqref{expo}, seen as a subset of ${\cal A}_{\text{mom}}$.

The issue, though, is that the group $G$ may have compact subgroups. As a consequence, the Lie algebra element $B(X,Y)$ in \eqref{baker} may be a nonzero logarithm of the identity: $\exp(B(X,Y))=e\in G$ even though $B(X,Y)\neq 0$. In order for exponential operators \eqref{expo} to yield a representation of $G$ when acting on a Hilbert space, one needs to constrain the operator algebra that is meant to be quantized in the first place. Namely, consider the normal subgroup $\mathcal{I}$ of $\mathcal{E}$ consisting of operators that act as the identity when exponential operators represent the group:
\begin{align}
\label{e9}
  \mathcal{I}
  :=
  \Big\{e^{-i\langle \hat p,X\rangle}\Big|X\in\mathfrak{g}\,\text{~such that }\exp(X) = e\Big\}\subset\cal E.
\end{align}
This is the `identity subset' of exponential operators. It is designed so that the group $G$ can be identified with the quotient $G\cong \mathcal{E}/\mathcal{I}$, in the same way that $\text{U(1)}\cong\mathbb{R}/\mathbb{Z}$.\footnote{Both $\mathcal{E}$ and $\mathcal{I}$ are groups, as defined around \eqref{eq: BCH}--\eqref{e9}, and $\mathcal{I}$ is a normal subgroup of $\mathcal{E}$, so the quotient $\mathcal{E}/\mathcal{I}$ is a group. In fact, it is the Lie group $G$ even though neither $\mathcal{E}$ nor $\mathcal{I}$ are Lie groups in general.} Its elements are the logarithms of the identity. One may think that this is the same as labelling the different branches of the logarithm from $\mathfrak{g}$ to $G$, but we will see that this is not so: the set \eqref{e9} typically overcounts these branches significantly.

\begin{figure}
\centering
\includegraphics[width=.8\textwidth]{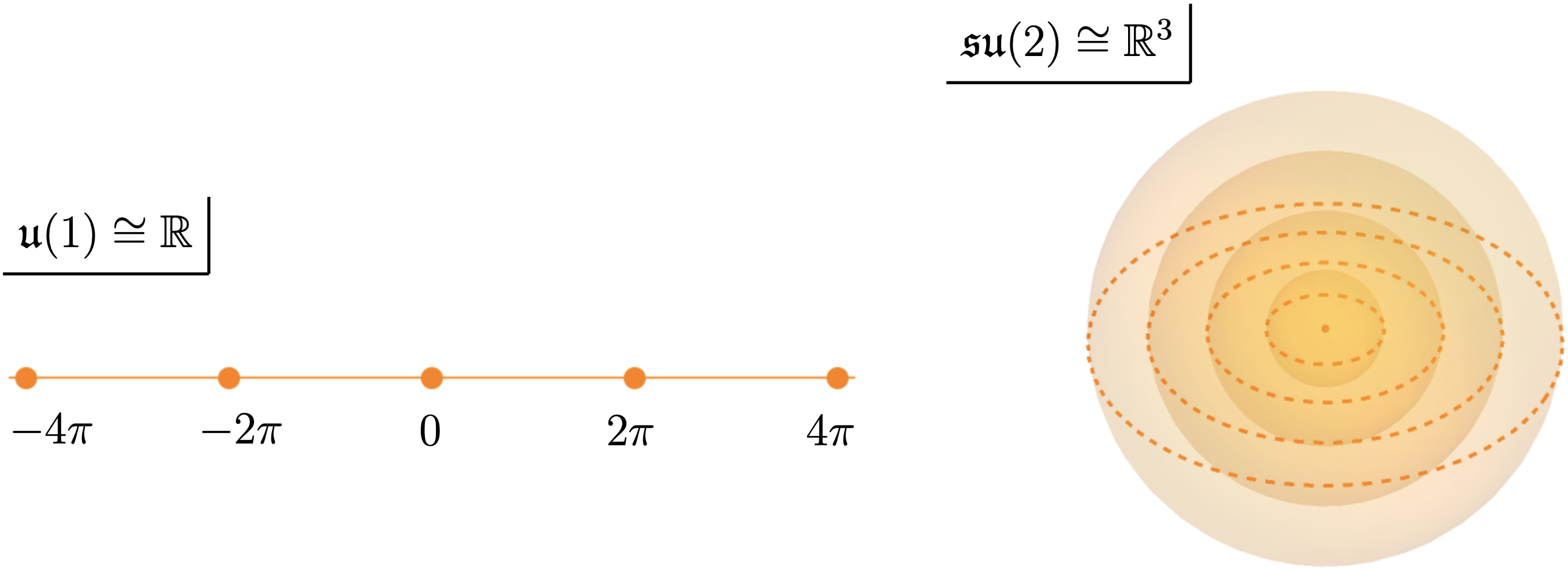}
    \caption{Logarithms of the identity in the $\mathfrak{u}(1)$ Lie algebra (left) and the $\mathfrak{su}(2)$ Lie algebra (right). In the U(1) case, logarithms of the identity form a one-dimensional lattice of $2\pi$-separated points in the Lie algebra $\mathfrak{u}(1)\cong\mathbb{R}$. This is because the exponential is $\exp: x\mapsto e^{ix}$. In the SU(2) case, we take the exponential map to be $\exp:\vec X \mapsto e^{i\vec X\cdot \vec \sigma}$ in terms of Pauli matrices (see section \ref{sec: examples}). With this convention, the logarithms of the identity in $\mathfrak{su}(2)\cong\mathbb{R}^3$ are concentric spheres with radii $2\pi n$, $n=0,1,2,...$, as predicted in general by eq.~\eqref{eq: sum over I unpacked}.}
    \label{fig:logs}
\end{figure}

The identity subgroup \eqref{e9} will appear extensively in this work, so let us describe it carefully. For any $g\in G$, let
\begin{equation}
\label{logg}
\operatorname{Logs}(g):=\{X\in \mathfrak{g}|\exp(X)=g\}\subset\mathfrak{g}
\end{equation}
be the set of its logarithms. Then one can write
\begin{equation}
\label{ix}
\mathcal{I}
\cong
\operatorname{Logs}(e)=\{X\in \mathfrak{g}| \exp(X)=e\}\subset\mathfrak{g},
\end{equation}
so characterizing $\cal I$ requires a good grasp on elements of $\mathfrak{g}$ that exponentiate to the identity. To this end, let $T\cong (\mathbb{S}^1)^r$ be a maximal torus of $G$, denote by $\mathfrak{t}\subset \mathfrak{g}$ its Lie algebra, and let $\{a_i|i=1,...,r\}$ be a basis of $\mathfrak{t}$ such that $\exp(2\pi a_i) = e$ but $\exp(\theta a_i)\neq e$ for all $\theta\in(0,2\pi)$. The elements of $\mathfrak{t}\cong \mathbb{R}^r$ that exponentiate to the identity are readily described in these terms: $X\in \mathfrak{t}$ exponentiates to the identity if and only if $X = 2\pi n^i a_i$ for some integers $n^1,...,n^r$.\footnote{Our convention for the exponential is $\mathbb{R}\to\mathbb{S}^1:x\mapsto e^{ix}$.} This provides the sought-for characterization of the identity subset \eqref{e9}--\eqref{ix}: since all maximal tori are mutually conjugate, one ends up with
\begin{align}
      \mathcal{I} 
      =
      \bigcup_{n^i\in \mathbb{Z}^r}\Big\{\mathrm{Ad}_{g}(2\pi n^i a_i)\,\Big|\,g\in G\Big\}
      = 
      \bigcup_{n^i\in \mathbb{Z}^r} 2\pi n^i \mathcal{O}_{a_i}\subset\mathfrak{g},
      \label{eq: sum over I unpacked}
\end{align}
where $\mathcal{O}_{a_i} := \{\mathrm{Ad}_g(a_i)|g\in G\}$ is the orbit of $a_i$ under the adjoint action, $\mathrm{Ad}_g(X):=gXg^{-1}$. Two examples of $\cal I$ are shown in fig.~\ref{fig:logs}.

By contrast, the set of logarithms \eqref{logg} of a generic group element $g\in G$ is discrete. This is because almost no element of $G$ is fixed by conjugation, in contrast to the identity; the only exceptions are elements in the center of $G$. Again, it will be essential for later purposes to describe the set \eqref{logg} explicitly, so let us do it here. Let $X\in \mathfrak{g}$ be a Lie algebra element that does not exponentiate to a central element in $G$, and let $T\cong(\mathbb{S}^1)^r$ be the maximal torus passing through $\exp(X)$. The Lie algebra $\mathfrak{t}$ of $T$ then contains $X$, so let $\{a_i(X)|i=1,\dots,r\}$ be a basis of $\mathfrak{t}$ such that $X\propto a_1(X)$, with $\exp(2\pi a_i(X))=e$ but $\exp(\theta a_i(X))\neq e$ for all $\theta \in (0,2\pi)$. Since $X$ is generic, all the logarithms of $\exp(X)$ can be expressed as discrete translations along the Lie algebra of $T$:
\begin{align}
    \operatorname{Logs}(\exp(X))=\{X+2\pi n^i a_i(X)\,|\, n^i\in \mathbb{Z}\}.
    \label{e84}
\end{align}
It follows that $|\operatorname{Logs}(X)| = |\mathbb{Z}|^r$ for almost any $X\in\mathfrak{g}$, and that each branch of the logarithm is labelled by an element of $\mathbb Z^r$. This will repeatedly be useful below; it ultimately ensures that all normalization issues encountered with nonabelian groups coincide with those one faces when seeing wave functions on a torus as periodic functions on $\mathbb{R}^n$.

\paragraph{Quotienting the operator algebra.} In order to ensure that eq.~\eqref{eq: BCH} represents the group $G$, which generally has compact subgroups, one needs to identify operators in $\cal A$ that differ from each other by the insertion of elements of the identity subset \eqref{e9} of exponential operators. This can be achieved by requiring that the algebra be blind to elements of $\mathcal{I}$, in the sense of identifying $\hat I \hat A$ and $\hat A \hat I$ for any operator $\hat A\in\cal A$ and any $\hat I\in\cal I$. In such a reduced algebra, any element in $\mathcal{I}$ commutes with all operators, and in that regard becomes (proportional to) the identity.

Let us again make this more precise, for it will be essential in understanding the difference between the two pairs of Fourier-dual representations built in section \ref{sec: representations}. Defining the two-sided ideal ${\cal A}_\mathcal{I}:= \big\langle\{\hat I \hat A\hat I ^{-1}-\hat A\,\big| \hat A\in {\cal A},~\hat I\in \mathcal{I}\}\big\rangle$ of $\cal A$, use it to quotient the operator algebra, thereby obtaining ${\cal A}/{\cal A}_{\cal I}$. The latter is isomorphic to the subalgebra of operators that are invariant under conjugation by $\mathcal{I}$. This is best seen by viewing the quotient ${\cal A}/{\cal A}_{\cal I}$ as isomorphic to the image of the projector ${\cal P}_\mathcal{I}$ that averages over conjugations by elements of $\mathcal{I}$:
\begin{align}
\label{e15}
       {\cal P}_\mathcal{I}(\hat A)
       :=
       \frac{1}{|{\cal I}|}\sum_{\hat I\in \mathcal{I}} \hat{I}\hat A\hat I ^{-1}.
\end{align}
Written in this way, the projector is ill-defined because $|{\cal I}|:=\operatorname{Vol}(\mathcal{I})$ is infinite. One can regularize the sum \eqref{e15} by introducing an `infrared' cutoff $\Lambda$, defining the set $\mathcal{I}_\Lambda := \{e^{-i\langle \hat p,X\rangle}| \,X \in \mathcal{I}\text{ such that }\|X\|\leq \Lambda\}$ where the norm is taken with respect to a metric of $\mathfrak{g}$, and defining a projector ${\cal P}_{\mathcal{I}_\Lambda}(\hat A):=\tfrac{1}{|\mathcal{I}_\Lambda|}\sum_{\hat I\in \mathcal{I}_\Lambda}\hat I \hat A\hat I ^{-1}$. Then the projector \eqref{e15} is the limit of ${\cal P}_{\mathcal{I}_\Lambda}$ as $\Lambda\to\infty$. Similar regularizations will appear repeatedly below, but they are to be expected when dealing with operators and Hilbert spaces on quotient manifolds, and they should be familiar to the reader from the relation between $L^2(\mathbb{R})$ and $L^2(S^1)$ (see section \ref{ssec: U(1) case}).

As a closing remark, note that the algebra ${\cal A}$ defined in eqs.~\eqref{5a} contains no information on the global structure of the group $G$, since coordinate functions are only locally defined. Taking the quotient $\mathcal A/\mathcal A_{\mathcal I}$ effectively reinstates the global structure of the group in the operator algebra. In fact, upon unraveling the definitions in the commutative U(1) case (see section \ref{sec: examples}), one finds that $\mathcal A$ is the operator algebra obtained by sums and products of $\hat X$s and $\hat P$s, while $\mathcal A/\mathcal A_\mathcal I$ is the subspace of `periodic' operators, which are invariant under translations by integer multiples of $2\pi$.

\section{Position and momentum representations}
\label{sec: representations}

In section \ref{sec: operator algebra}, we introduced the abstract operator algebra $\mathcal A$ obtained by putting hats on the coordinates of phase space $G\times \mathfrak{g}^*$. This algebra bears, by construction, only local information. In order to reinstate the global properties of the group, we projected the operator algebra down to $\mathcal A/ \mathcal{A}_{\mathcal I}$.

Our goal now is to represent both algebras on Hilbert spaces by defining, for each of them, a position representation and a momentum representation. The position representation of $\mathcal A/\mathcal A_\mathcal I$ is defined in section \ref{ssec: position rep 1} and turns out to be the regular representation of the group on the Hilbert space $L^2(G)$. As we will see in section \ref{ssec: position rep 2}, the latter can be embedded in the larger space of functions on the Lie algebra, $L^2(\mathfrak{g})$. This mimics the fact that functions on U(1) can be seen as the subset of periodic functions inside the set of functions on the real line $\mathbb R = \mathfrak{u}(1)$; the larger space $L^2(\mathfrak g)$ is, similarly, the position representation of the larger algebra $\mathcal A$. As far as momentum representations are concerned, we define in section \ref{ssec: momentum rep 1} the momentum representation of $\mathcal A$, sometimes denoted $L^2_\star(\mathfrak g^*)$ in earlier literature (see \textit{e.g.}~\cite{Guedes:2013vi}). That part is the noncommutative one: it generalizes the notion of momentum space to the nonabelian case, consisting of noncommutative wave functions of continuous momenta. Finally, section \ref{ssec: momentum rep 2} projects the momentum representation of $\mathcal{A}$ down to that of the quotient $\mathcal{A}/\mathcal A_\mathcal I$, literally by projection onto the $\mathcal{I}$-invariant subspace. This projection generalizes, for nonabelian groups, the passage from functions of continuous momenta in $L^2(\mathbb R)$ to Fourier coefficients in $\ell^2(\mathbb Z)$.

\subsection{Small position representation}
\label{ssec: position rep 1}

We start by representing the operator algebra $\cal A$ in the most straightforward way, on the Hilbert space $L^2(G)$ of square-integrable wave functions on $G$ \cite{Mukunda:2003wna}. This Hilbert space is actually obtained by geometric quantization of the cotangent bundle $T^*G$ in real polarization \cite{woodhouse1997}.

\paragraph{Hilbert space $\boldsymbol L^{\boldsymbol2}\boldsymbol(\boldsymbol G\boldsymbol)$.} We denote wave functions on $G$ by lowercase Greek letters $\phi$, $\psi$, etc. Their scalar product is
\begin{align}
\label{e18}
      \langle \phi|\psi\rangle_G 
      :=
      \int_G \dd g\; \bar\phi(g)\psi(g) 
      =
      \!\!\!\!\int\limits_{\substack{\text{principal}\\ \text{branch in $\mathfrak{g}$}}} \!\!\!\!J(X)\dd^n X\,\bar\phi(X)\psi(X),
\end{align}
where $\dd g$ is the left Haar measure on $G$ while $\dd^nX$ is the Lebesgue measure on $\mathfrak{g}\cong\mathbb{R}^n$ in principal branch coordinates defined above \eqref{eq: prop of pb coordinates}. We write $J(X)$ for the corresponding Jacobian of the Haar measure, namely \cite{Taylor} (see also \cite{Helgason})
\begin{align}
\label{e19}
     J(X)
     =
     \det\left(\frac{1-e^{-\mathrm{ad}_X}}{\mathrm{ad}_{X}}\right),
\end{align}
where $\mathrm{ad}_X:=[X,\cdot\,]$ is the adjoint action by $X\in\mathfrak{g}$. This Jacobian will appear repeatedly below, anytime an integral over $G$ is expressed in principal branch coordinates. If $G$ is unimodular (left and right Haar measures coincide), one has $\operatorname{tr}(\mathrm{ad}_X)=0$ and eq.~\eqref{e19} can be recast as $J(X)=\frac{\sinh(\mathrm{ad}_X/2)}{\frac{1}{2}\mathrm{ad}_X}$, in which case the Jacobian is an even function of $X$. We will exploit this in section \ref{sec: examples} to derive nonabelian Poisson summation formulas.

\paragraph{Regular representation of $\boldsymbol G$.} The action of position and momentum operators on a wave function is now defined by the \i{position representation} that stems from the left regular representation of $G$:
\begin{align}
\label{e20}
     (\hat X^j\psi)(g)
     :=
      X^j(g)\psi(g),
      \qquad
      (\hat p_j \psi)(g)
      :=
      -i(\mathcal{L}_j \psi)(g).
\end{align}
This enforces the commutation relations \eqref{5a} with operators $\hat X^i$, $\hat p_i$ that are Hermitian, as they should be. Strictly speaking, the action of position operators $\hat X^j$ in \eqref{e20} is ill-defined because the coordinates $X^j(g)$ only make sense locally. However, the action of any `potential operator' $\hat V={\cal Q}(V(g))$ is well defined and reads $(\hat V\psi)(g):=V(g)\psi(g)$, where $V(g)$ is any smooth function on $G$.

As for momenta, having $\hat p_i$ act as a derivation along $t_i$ in \eqref{e20} ensures that its exponential acts by the flow of the vector field $t_i$, \i{i.e.}~by left multiplication of the argument of a wave function. This is indeed to say that the second equation in \eqref{e20} is the infinitesimal counterpart of the left regular representation of $G$. As a result, exponential operators \eqref{expo} yield a representation \eqref{eq: BCH} of $G$ via $(e^{-i\langle \hat p,Y\rangle}\cdot\psi)(g)=(e^{-\mathcal{L}_Y}\psi)(g)=\psi(\exp(Y)g)$. It follows that the identity subset \eqref{ix} acts trivially in this representation, \i{i.e.}
\begin{equation}
\label{s8}
     (e^{-i\langle \hat p,Y\rangle}\psi)(g)=\psi(\exp(-Y)g)=\psi(eg)=\psi(g)
\end{equation}
whenever $\exp(Y)=e$. Any wave function is thus trivially $\cal I$-invariant. In other words, it is really the quotient space ${\cal A}/{\cal A}_{\cal I}$ that acts in the position representation \eqref{e20}. As we will see in section \ref{ssec: position rep 2}, the projection \eqref{e15} of position operators to $P_\mathcal{\mathcal I}\cdot\mathcal A_\text{pos}\subset \mathcal A/\mathcal A_{\mathcal I}$ leads to \i{bona fide} potential operators that act in a well-defined way, in contrast to \i{e.g.}~$\hat X^j\in \mathcal A$.

In most cases below, it will be clear that $\hat A \psi(g)$ denotes the action of $\hat A\in {\cal A}/{\cal A}_{\cal I}$ on the function $\psi\in L^2(G)$ in the position representation. In places where confusion may occur, we will explicitly denote the representation map by $\rho_\text{pos}^\mathcal{I}: {\cal A}/{\cal A}_{\cal I}\rightarrow \operatorname{End}(L^2(G))$, meaning that $(\hat A \psi)(g) := (\rho_\text{pos}^\mathcal{I}(\hat A)\psi)(g)$. The $\mathcal{I}$ superscript is added here to stress that one is representing the projected operator algebra $\mathcal A/\mathcal A_{\mathcal I}$. By contrast, we will define in section \ref{ssec: position rep 2} a `larger' position representation $\rho_{\text{pos}}$, without superscript, where the full algebra $\mathcal A$ acts on wave functions.

\subsection{Large momentum representation and star product}
\label{ssec: momentum rep 1}

Here we introduce an action of position and momentum operators on wave functions that live on the entire momentum space $\mathfrak{g}^*$, \i{i.e.}~on the dual of the Lie algebra of $G$. In particular, this will involve a noncommutative action of momenta. It will quickly be apparent that the resulting Hilbert space is much larger than $L^2(G)$, so modifications will be needed to relate the two. These are addressed separately in sections \ref{ssec: momentum rep 2}--\ref{ssec: position rep 2}.

\paragraph{Momentum operators.} The momentum representation of the operator algebra $\cal A$ should act on functions $\Phi(p)$, $\Psi(p)$, etc.~that depend on momentum $p\in\mathfrak{g}^*$. However, naively defining  $(\hat p_i \Phi)(p)=p_i\Phi(p)$ fails to reproduce the last commutator in eqs.~\eqref{5a}, so one needs to deform the multiplication law on the space of functions on $\mathfrak{g}^*$ \cite{Oriti:2011ac,Raasakka:2011np,Guedes:2013vi}. This is achieved  by a star product $\star$ that replaces pointwise multiplication, such that
\begin{align}
     \hat p_i \Phi(p)
     =
     p_i\star\Phi(p)\qquad\text{and}\qquad
     p_i\star p_j-p_j\star p_i =- i  c_{ij}{}^k p_k.
     \label{eq: momentum op in momentum space}
\end{align}
Not all ambiguities in the definition of the star product are fixed by the last commutators in eqs.~\eqref{5a}, but those ambiguities that remain are fixed by the choice of quantization map \eqref{eq: quantization map}. Indeed, let $\hat A= \mathcal{Q}(A)$ be the quantization of a function $A(p)$; then the action of $\hat A$ on a function $\Phi$ in the momentum representation is given by $(\hat A\Phi)(p) = A(p)\star \Phi(p)$.

The star product of functions in momentum space is thus fixed by one's choice of momentum quantization scheme. As in the Moyal case \cite{Zachos}, it reads
\begin{align}
    A\star B 
    :=
    \mathcal{Q}_{\text{mom}}^{-1}(\hat A \hat B),\qquad \forall A,B\in C^\infty(\mathfrak{g}^*),
    \label{eq: def star product}
\end{align}
where $\mathcal{Q}_{\text{mom}}$ is the quantization map \eqref{eq: quantization map} restricted to functions that only depend on momenta. For the symmetric ordering \eqref{eq: symmetric ordering}, eq.~\eqref{eq: def star product} defines the so-called Gutt star product on $\mathfrak{g}^*$ \cite{Gutt}. One readily deduces from \eqref{eq: def star product} that the star product is well-behaved under conjugation:
\begin{align}
       \overline{A\star B}
       =
       \bar{B}\star \bar{A}.
       \label{eq: prop star under conjugation}
\end{align}
To avoid confusion, we will sometimes stress the variable of the star product by a subscript, as in \i{e.g.}~$\Phi(p)\star \Psi(p)=\Phi(p)\star_p\Psi(p)$. 

Using the fact that exponential functions are quantized to exponential operators \eqref{expo} and the Baker-Campbell-Hausdorff property \eqref{eq: BCH} of momentum operators, star products of plane waves read 
\begin{align}
e^{-i\langle p,X\rangle}\star e^{-i\langle p, Y\rangle} 
=
e^{-i\langle p, B(X,Y)\rangle}.
\label{e26}
\end{align}
This feature is the defining property of the Gutt star product \cite{Gutt}. Indeed, since plane waves provide a basis of the space of functions of $p$, knowing how the star product behaves on them suffices to compute the star product of {\it any} two functions. Specifically, the bidifferential operator implementing the star product can be written as $\Phi\star \Psi(p) = \Phi(p) e^{-i\langle p, B(i\overleftarrow \partial, i\overrightarrow\partial)-i \overleftarrow\partial - i \overrightarrow\partial\rangle} \Psi(p)$ for any two smooth functions $\Phi,\Psi$ in momentum space \cite{Dito}.

\paragraph{Position operators.} Position operators need to commute and to be canonically conjugate to momentum operators, as in eqs.~\eqref{5a}. As in standard quantum mechanics, one may define
\begin{align}
       (\hat X^i\Phi)(p)
       :=
       i \frac{\partial\Phi}{\partial p_i}(p).
       \label{eq: zeta momentum rep}
\end{align}
\noindent\i{Statement.} The \i{momentum representation} of $\cal A$ defined in eqs.~\eqref{eq: momentum op in momentum space} and \eqref{eq: zeta momentum rep} reproduces the commutators \eqref{5a}.
\begin{center}
\begin{minipage}{.9\textwidth}
{\footnotesize%
\i{Proof.} The vanishing commutator $[\hat X^i,\hat X^j]=0$ trivially holds by the definition \eqref{eq: zeta momentum rep}, and the star product in \eqref{eq: momentum op in momentum space} guarantees that the last commutator of eqs.~\eqref{5a} holds as well. Thus, it suffices to focus on the second commutator in eqs.~\eqref{5a}, namely $[\hat X^j,\hat p_i]=i \widehat{\mathcal{L}_i X^j}$. The latter is true, as an operator acting on all functions of $p$, if and only if it holds true on exponentials $e^{-i\langle p,X\rangle}$, which form a basis of the space of functions of $p$. Let us therefore focus on computing the commutator $[\hat X^i,\hat p_j]$ acting on exponentials, namely
\begin{align*}
[ \hat X^j,\hat p_i]e^{-i\langle p,Y\rangle} 
&= i\frac{\partial}{\partial p_j}\big(p_i\star e^{-i\langle p,Y\rangle}\big) - p_i\star Y^j e^{-i\langle p,Y\rangle}\\
&=
-\frac{\partial}{\partial p_j}\frac{\partial}{\partial Z^i}\bigg|_{Z=0}e^{-i\langle p,Z\rangle}\star e^{-i\langle p,Y\rangle}
-i Y^j \frac{\partial}{\partial Z^i}\bigg|_{Z=0} e^{-i\langle p,Z\rangle}\star e^{-i\langle p,Y\rangle}
\end{align*}
where we used $\hat X^j= Y^j$ when acting through \eqref{eq: zeta momentum rep} on a plane wave $e^{-i\langle p,Y\rangle}$. By virtue of the star product \eqref{e26}, one now has
\begin{align*}
[ \hat X^j,\hat p_i]e^{-i\langle p,Y\rangle} 
&=
i\frac{\partial}{\partial Z^i}\bigg|_{Z=0} \left(B(Z,Y)^j e^{-i\langle p,B(Z,Y)\rangle}\right)-i Y^j \frac{\partial}{\partial Z^i}\bigg|_{Z=0} e^{-i\langle p,B(Z,Y)\rangle}\\
&=
i\left(\frac{\partial}{\partial Z^i}\bigg|_{Z=0} B(Z,Y)^j\right) e^{-i\langle p,B(Z,Y)\rangle}.
\end{align*}
Finally, since the Lie derivative along $t_i\in\mathfrak{g}$ is an infinitesimal translation along the $i^{\text{th}}$ direction of the $Y$ vector, eq.~\eqref{eq: prop of pb coordinates} yields $i \widehat{\mathcal L_i X^j} = i\frac{\partial}{\partial Z^i}\big|_{Z=0}B(Z,Y)^j$ when acting on a plane wave $e^{-i\langle p,Y\rangle}$, from which it follows that $[ \hat X^j,\hat p_i]e^{-i\langle p,Y\rangle}=i \widehat{\mathcal L_iX^j} e^{-i\langle p,Y\rangle}$.\hfill$\blacksquare$}
\end{minipage}
\end{center}

\paragraph{Noncommutative scalar products.} Having defined the action \eqref{eq: momentum op in momentum space}--\eqref{eq: zeta momentum rep} of position and momentum operators, the missing ingredient is the scalar product of wave functions. The carrier space of the momentum representation is the Hilbert space of $\star$-square-integrable functions on $\mathfrak{g}^*$, denoted $L^2_\star(\mathfrak{g}^*)$, with scalar product
\begin{align}
     \langle\Phi|\Psi\rangle_{\mathfrak{g}^*}
     :=
      \int_{\mathfrak{g^*}}\frac{\dd^n p}{(2\pi )^n}\;(\bar\Phi\star\Psi)(p),
     \label{eq: scalar product momentum space}
\end{align}
where $\dd^np$ is the Lebesgue measure on $\mathfrak{g}^*$ and the star product is that introduced around \eqref{eq: def star product}. This bilinear form is sesquilinear thanks to \eqref{eq: prop star under conjugation}. It also satisfies the key property of being invariant under exponential operators \eqref{expo}, which is to say that the set $\cal E$ of such operators acts unitarily. Indeed, for any plane wave $e^{-i\langle \hat p,X\rangle}\in \cal E$, one has
\begin{equation}
\begin{split}
       \big\langle e^{-i\langle \hat p,X\rangle} \Phi\big|e^{-i\langle \hat p,X\rangle} \Psi\big\rangle_{\mathfrak{g}^*}
       &=
       \int _{\mathfrak{g}^*}\frac{\dd^n p}{(2\pi)^{n}}\; \bar\Phi \star e^{i\langle p, X\rangle}\star e^{-i\langle p,X\rangle}\star \Psi\\
     & =
     \int _{\mathfrak{g}^*}\frac{\dd^n p}{(2\pi)^{n}}\; \bar\Phi \star e^{-i\langle p,B(-X,X)\rangle}\star \Psi
      =\langle\Phi|\Psi\rangle_{\mathfrak{g}^*}
      \end{split}
      \label{eq: momentum scalar product E inv}
\end{equation}
owing to \eqref{e26} and $B(X,-X)=0$ by the Baker-Campbell-Hausdorff formula \eqref{baker}.

A potential issue at this point is that the pairing \eqref{eq: scalar product momentum space} is not manifestly positive-definite. This will be remedied in section \ref{sec:DefFourier} thanks to the noncommutative Fourier transform: the latter will provide an isometry between $L^2_\star(\mathfrak{g}^*)$ and the space $L^2(\mathfrak{g})$ of square-integrable wave functions on the Lie algebra, endowed with a \i{bona fide} scalar product and closely related to $L^2(G)$.

Finally, a remark on notation. We mostly avoid explicitly writing the representation map but, when a specification is needed, we denote it as $\rho_{\text{mom}} : {\cal A}\rightarrow \operatorname{End}(L^2_\star(\mathfrak{g}^*))$, meaning that $(\hat A \Phi)(p):= (\rho_{\text{mom}}(\hat A)\Phi)(p)$. This should be compared with the notation $\rho^\mathcal{I}_{\text{pos}}$ introduced at the end of section \ref{ssec: position rep 1}. In particular, note that there is no superscript in the momentum representation defined so far, for reasons that we now explain.

\paragraph{The problem.} The momentum representation, as built so far, is not isomorphic to the position representation \eqref{e20}. Indeed, in contrast to eq.~\eqref{s8}, the action of the identity subset \eqref{e9} on $L^2_{\star}(\mathfrak{g}^*)$ is nontrivial since, in general,
\begin{equation}
\label{e38}
e^{-i\langle \hat p, Y\rangle}\star \Phi(p)\neq \Phi(p)\text{ even when }\exp(Y)=e.
\end{equation}
One might have hoped that star products save the day, but they do not, as the issue is one of topology rather than commutativity: even in the simplest case of the (commutative) U(1) group, with a pointwise (commutative) star product of functions on $\mathfrak{u}(1)^*=\mathbb{R}$, the identity subset $\mathcal{I}=\{e^{-2\pi inp}\,|\,n\in\mathbb{Z}\}$ is such that $e^{-2\pi inp}\Phi(p)\neq \Phi(p)$ in general. The problem, in other words, lies in the fact that the group $G$ may have compact subgroups. The result is that the momentum representation on $L^2_{\star}(\mathfrak{g}^*)$ is `larger' than the position representation on $L^2(G)$. To retrieve an isomorphism between them, one is left with two choices: either make the momentum space representation smaller, or make the position space representation larger. These are respectively treated in sections \ref{ssec: momentum rep 2} and \ref{ssec: position rep 2}.

\subsection{Small momentum representation}
\label{ssec: momentum rep 2}

What spoils the isomorphism between momentum and position representations is summarized in \eqref{e38}: the fact that states in $L^2_\star(\mathfrak{g}^*)$ are typically not $\mathcal{I}$-invariant, where $\cal I$ is the identity subset \eqref{e9} or \eqref{ix}. A straightforward solution is to project the large space $L^2_{\star}(\mathfrak{g}^*)$ on its subspace consisting of $\cal I$-invariant functions. Analogously to \eqref{e15}, this is achieved by the projection operator
\begin{align}
\label{proji}
\hat P_\mathcal{I}
:=
\frac{1}{|{\cal I}|}\sum_{\hat I\in \mathcal{I}}\hat I
=
\frac{1}{|{\cal I}|}\sum_{Y\in \operatorname{Logs}(e)}
    e^{-i\langle\hat p,Y\rangle},
\end{align}
whereupon the image $\Im(\hat P_\mathcal{I})=\hat P_\mathcal{I}\cdot L^2_{\star}(\mathfrak{g}^*)$ is the space of $\cal I$-invariant functions. The scalar product on that space is the restriction of \eqref{eq: scalar product momentum space} to $\cal I$-invariant functions.

Because $|{\cal I}|$ is typically infinite, acting with $\hat P_\mathcal{I}$ on finite-normed wave functions in $L^2_\star(\mathfrak{g}^*)$ produces wave functions of vanishing norm. The same subtlety affected averaged operators \eqref{e15}, and a similar regularization is available for the operator \eqref{proji}. This is actually a minor problem, identical to the one relating Fourier \i{transforms} and Fourier \i{series}, \i{i.e.}~functions on $\mathbb{R}$ and functions on $S^1$ (see section \ref{sec: examples}). The way out is to allow oneself to renormalize states after projection with infinite factors $\sqrt{|\mathbb Z|^r}$ for some $r$, in order to get a result that is finite in norm. This is analogous to infrared regularization in standard quantum mechanics: one works in finite volume with periodic boundary conditions, before taking the thermodynamic limit at the end of the day.

Note that the operator algebra which acts on the $\cal I$-invariant Hilbert space $\hat P_\mathcal{I}\cdot L^2_{\star}(\mathfrak{g}^*)$ is not quite $\cal A$, but its quotient ${\cal A}/{\cal A}_{\cal I}$ introduced around \eqref{e15}. Equivalently, one may represent operators $\hat A\in {\cal A}$ via the $\cal I$-invariant momentum representation $\rho_{\text{mom}}^\mathcal{I}$ defined by
\begin{align}
\label{e40}
\rho_{\text{mom}}^\mathcal{I}(\hat A)
:=
 \hat P_\mathcal{I}\rho_{\text{mom}}(\hat A)\hat P_\mathcal{I},
\end{align}
where $\rho_{\text{mom}}$ is the `large' momentum representation of section \ref{ssec: momentum rep 1} and $P_\mathcal{I}$ is the projector \eqref{proji}. We will use Fourier transforms in section \ref{sec:DefFourier} to show that the $\cal I$-invariant momentum representation \eqref{e40} is unitarily equivalent to the position representation \eqref{e20} on $L^2(G)$. At first sight, convergence issues could arise when projecting $L^2_\star(\mathfrak g^*)$ through $\hat P_\mathcal I$. We show in section \ref{ssec: noncommutative Fourier coefficients} that this is not the case, by providing a basis of $\hat P_\mathcal{I}\cdot L^2_\star(\mathfrak g^*)$.

\paragraph{Comparison with \cite{Oriti:2011ac,Raasakka:2011np,Guedes:2013vi}.} We stress that our current construction differs from that of \cite{Oriti:2011ac,Raasakka:2011np,Guedes:2013vi}, as follows. First, note that the vector space $\hat P_\mathcal{I}\cdot L^2_\star(\mathfrak{g}^*)$ is isomorphic to the vector space of coinvariants under the action of $\mathcal{I}$:
\begin{align}
\label{quop}
     \hat P_\mathcal{I}\cdot L^2_\star(\mathfrak{g}^*)
     \cong
     L^2_\star(\mathfrak{g}^*)\Big/\big\langle\{\hat I \Phi - \Phi\,|\,\hat I \in \mathcal{I},~\Phi\in L^2_\star(\mathfrak{g}^*)\}\big\rangle.
\end{align}
This is readily shown by representing a class of functions on the right-hand side by the unique $\mathcal{I}$-invariant function in that class. In ref.~\cite{Oriti:2011ac,Raasakka:2011np,Guedes:2013vi}, it is the quotient space on the right-hand side of \eqref{quop} that is chosen as carrier space for the momentum representation.

The issue is that defining a scalar product on that quotient space cannot be done consistently without averaging over $\mathcal{I}$. Indeed, any two representatives of the class $[\Phi]$ differ by the action of some $\hat I \in \mathcal{I}$; if one attempts to define the scalar product as in \cite{Oriti:2011ac,Raasakka:2011np,Guedes:2013vi}, by $\langle [\Phi]|[\Psi]\rangle := \int_{\mathfrak{g}^*}{\dd^n p}\,\bar \Phi(p)\star\Psi(p)$, then a key consistency check is that the scalar product must not depend on the choice of representatives for $[\Phi]$ and $[\Psi]$. The problem is that it \i{does} depend on that choice: picking another representative for $[\Phi]$, say $e^{-i\langle p,Y\rangle}\star\Phi(p)$ with $\exp(Y)=e$, one generally has $\int\dd^n p\,\bar\Phi(p)\star\Psi(p)\neq\int\dd^np\,\bar\Phi(p)\star e^{i\langle p ,Y\rangle}\star \Psi(p)$. This confirms that the only way to properly define the scalar product of equivalence classes is to average the scalar product of representatives over whole classes. This, in turn, amounts to working in the subspace $\hat P_\mathcal{I}\cdot L^2_\star(\mathfrak{g}^*)$ of $\cal I$-invariant wave functions, as done here. A similar subtlety affects the definition of operators acting on wave functions in momentum space.

\subsection{Large position representation}
\label{ssec: position rep 2}

Let us now turn to the second way of relating position and momentum representations, by making the position representation of section \ref{ssec: position rep 1} suitably `larger'. We begin with the simple example of the U(1) group to illustrate the procedure, then apply it to more general Lie groups.

\paragraph{A simple example: functions on $\mathbb{R}\hspace{-.69em}\mathbb{R}$.} We saw below \eqref{e38} that there is no isometry between $L^2(\text{U(1)})$ and $L^2_\star(\mathfrak{u}(1)^*)=L^2(\mathbb{R})$ because the action of `identity operators' \eqref{e9} is nontrivial. In the case $G=\text{U}(1)$, the set of identity operators is $\mathcal{I} = \{e^{-2i\pi n \hat p}\,|\; n\in \mathbb{Z}\}$, which acts on $L^2_\star(\mathfrak{u}(1)^*)=L^2(\mathbb{R})$ according to
\begin{align}
     e^{-2\pi in\hat p}\Phi(p)
     =
     e^{-2\pi inp}\Phi(p).
     \label{eq: example blow up momentum}
\end{align}
This should be contrasted with the trivial action of identity operators on $L^2(U(1))$:
\begin{align}
     e^{-2\pi in\hat p}\phi(x)=\phi(x - 2\pi n) = \phi(x).
     \label{eq: example blow up position}
\end{align}
How to reconcile these two statements? Had we not assumed that the function $\phi$ is $2\pi$-periodic, eq.~\eqref{eq: example blow up momentum} would have been the Fourier transform of \eqref{eq: example blow up position} upon identifying $\Phi$ with the Fourier transform of $\phi$. In this respect, the space isomorphic to $L^2_\star(\mathfrak{u}(1)^*)$ is in fact $L^2(\mathfrak{u}(1))= L^2(\mathbb{R})$, and $L^2(U(1))$ is its subset consisting of periodic functions. This simple fact can be generalized to any Lie group.

Before turning to the generic case, let us discuss how the isometry between $L^2(U(1))$ and periodic functions in $L^2(\mathbb{R})$ is to be understood. The main objection is that nonzero periodic functions on $\mathbb{R}$ are never square-integrable, since they do not decay at infinity. This is the usual problem of plane waves in quantum mechanics, which technically belong to a `rigged' Hilbert space rather than $L^2(\mathbb{R})$ \cite{ballentine2014}. Formally, a solution is to keep track of divergent normalization factors such as $1/|\mathbb{Z}|$ in order to get finite results. This is easily seen by picking, at random, a function $\phi_0\in L^2(\mathbb{R})$ that decreases sufficiently fast at infinity. Applying the projector \eqref{proji} to $\phi_0$ yields
\begin{align}
\label{e46}
     \hat P_\mathcal{I}\phi_0(x) 
     = \frac{1}{|\mathbb{Z}|}\sum_{n\in \mathbb{Z}}\phi_0(x+2\pi n) 
     =: \frac{1}{|\mathbb{Z}|}\phi(x),
\end{align}
where $\phi(x)$ is now $2\pi$-periodic. As no function in $L^2(\mathbb{R})$ is $\mathcal{I}$-invariant (here meaning $2\pi$-periodic), the right-hand side of \eqref{e46} vanishes. This is confirmed by computing the norm of $\phi$:
\begin{align}
     \|\phi\|_\mathbb{R}^2 = \frac{1}{|\mathbb{Z}|^2}\int_\mathbb{R} \dd x \,|\phi(x)|^2
     =\frac{|\mathbb{Z}|}{|\mathbb{Z}|^2}\int_{-\pi}^\pi \dd x\,|\phi(x)|^2
     =\frac{1}{|\mathbb{Z}|}\|\phi\|^2_{U(1)}.
\end{align}
In order to get a result that is finite in norm, one thus needs to multiply $\hat P_\mathcal{I}\psi_0$ by a factor $\sqrt{|\mathbb{Z}|}$. This is to say that, for periodic functions $\phi$, $\psi$, one has
\begin{equation}
\label{e48}
\Big< \frac{1}{\sqrt{|\mathbb{Z}|}}\phi\Big|\frac{1}{\sqrt{|\mathbb{Z}|}}\psi\Big>_{\mathbb{R}} 
=
 \langle \phi|\psi\rangle_{\text{U(1)}}
\end{equation}
where both sides are now understood to be finite, generally nonzero. In all such cases, one should view the factor $|\mathbb{Z}|$ as specifying a prescription for finite-volume regularization.

\paragraph{General case: functions on $\mathfrak{g}\!\!\!\mathfrak{g}$.} In order to adapt the discussion to any exponential Lie group, the idea is to unwrap $G$ onto its Lie algebra and define the space $L^2(\mathfrak{g})$ of square-integrable functions on $\mathfrak{g}$, as follows. First define the \i{Baker-Campbell-Hausdorff group} as being the Lie algebra $\mathfrak{g}$, endowed with a group law \begin{equation}
X\cdot Y:=B(X,Y)
\label{xy}
\end{equation}
given by the Baker-Campbell-Hausdorff expansion \eqref{baker}. The identity is just $0$ and the inverse of an element $X$ is given by $X ^{-1}=-X$. Then consider the Hilbert space $L^2(\mathfrak{g})$, with the scalar product on the right-hand side of \eqref{e18} extended to the whole Lie algebra:
\begin{align}
     \langle \phi|\psi\rangle_{\mathfrak{g}}
     :=
     \int_\mathfrak{g}J(X)\dd^n X\; \bar \phi(X)\psi(X).
\end{align}
The action of position and momentum operators is defined by \eqref{e20}, as in $L^2(G)$. In particular, exponential operators act on that Hilbert space by (noncommutative) translations
\begin{align}
    e^{-i\langle \hat p, Y\rangle}\cdot\phi(X)
    =
    \phi(B(-Y,X)).
    \label{big nctranslation}
\end{align}
The action of `identity operators' \eqref{e9} is thus nontrivial, since in general $B(-Y,X)\neq X$ even when $\exp(Y)=e$. Also note that the action of position operators $\hat X^j$ as multiplication by $X^j$ is now well-defined, so that indeed the full operator algebra $\mathcal A$ is represented on that space. Let us therefore denote this alternative position representation $\rho_\text{pos}$, this time with no $\mathcal{I}$ superscript. We will see in section \ref{sec:DefFourier} that $L^2(\mathfrak{g})$ is isometric to $L^2_\star(\mathfrak{g}^*)$ through the noncommutative Fourier transform.

Similarly to the U(1) case, functions on $G$ can be seen as functions on $\mathfrak{g}$ that are periodic under the action of $\mathcal{I}$. Such functions take the same value at all logarithms of any given group element: $C^\infty(G)\cong \{\phi\in C^\infty(\mathfrak{g})\,|\,\phi(X)=\phi(B(Y,X))~~\forall X\in\mathfrak{g},~\forall Y\in{\cal I}\}$. One can again formally get an isometry between $L^2(G)$ and periodic functions in $L^2(\mathfrak{g})$, by defining $L^2(G)=\Im(\hat P_\mathcal{I})$ in terms of the projector \eqref{proji}. From the same reasoning as in the U(1) case \eqref{e48}, periodic functions in $L^2(\mathfrak{g})$ that are finite in norm are obtained by renormalizing states in the image of $\hat P_\mathcal{I}$ by a factor of $\sqrt{|\mathbb Z^r|}$. As a result, for $\phi,\psi$ that are $\cal I$-periodic in $C^\infty(\mathfrak{g})$, the analogue of eq.~\eqref{e48} reads
\begin{equation}
\begin{split}
     \bigg\langle \frac{1}{\sqrt{|\mathbb{Z}|^r}}\phi\bigg|\frac{1}{\sqrt{|\mathbb{Z}|^r}}\psi\bigg\rangle_{\!\!\mathfrak{g}} 
     &= \frac{1}{|\mathbb Z|^r}\int_\mathfrak{g}J(X)\dd^n X\; \bar \phi(X)\psi(X) 
     \\
     &= \!\!\!\!\int\limits_{\substack{\text{principal}\\ \text{branch in $\mathfrak{g}$}}} \!\!\!\!J(X)\dd^n X\; \bar \phi(X)\psi(X)
     =\langle\phi|\psi\rangle_G,
     \end{split}
     \label{e5354}
\end{equation}
where the principal branch of the logarithm was defined around eq.~\eqref{eq: prop of pb coordinates} and where we used that the branches of logarithm are labelled by $\mathbb Z^r$ as explained below \eqref{e84}. The bottom line here is the scalar product \eqref{e18} in $L^2(G)$, which was the expected result.

To conclude, we have now represented the operator algebras ${\cal A}$ and ${\cal A}/{\cal A}_{\cal I}$ on two pairs of Hilbert spaces. These were $L^2(G)\cong \hat P_{\cal I}\cdot L^2(\mathfrak{g})\subset L^2(\mathfrak{g})$ in position space, and $\hat P_{\cal I}\cdot L^2_\star(\mathfrak{g}^*)\subset L^2_\star(\mathfrak{g}^*)$ in momentum space. The same projector \eqref{proji} appears in both cases, by construction. We will now define intertwiners linking these representations two by two, namely a noncommutative Fourier transform and its inverse for the `large' representations, and Fourier coefficients and Fourier series for the `small' representations.

\section{Noncommutative Fourier transforms and series}
\label{sec:DefFourier}

The passage from position to momentum representations of $\cal A$ is achieved by a unitary intertwining operator, \i{i.e.}~an isometry that commutes with the action of operators. The isometry is just a Fourier transform in standard quantum mechanics. By analogy, following \cite{Oriti:2011ac,Raasakka:2011np,Guedes:2013vi}, we now refer to it as the \i{noncommutative Fourier transform} to stress that momenta fail to commute as in eqs.~\eqref{5a}. (It is also sometimes called the \i{group Fourier transform} \cite{Freidel:2005bb,Joung:2008mr}.)

Being an intertwiner is a stronger requirement than being a Hilbert space isomorphism. In our case, if $\cF$ is the sought-for intertwiner between `large' position and momentum representations, then the following two diagrams need to commute for any operator $\hat A$:
\begin{equation}
     \begin{tikzcd}
        L^2(\mathfrak{g})
        \arrow{r}{\cF}
        \arrow{d}{\rho_{\text{pos}}(\hat A)}
        &
        L^2_\star(\mathfrak{g}^*)
        \arrow{d}{\rho_{\text{mom}}(\hat A)}
        \\
        L^2(\mathfrak{g})
        \arrow{r}{\cF}
        &
        L^2_\star(\mathfrak{g}^*)
     \end{tikzcd}
     \qquad\text{and}\qquad
     \begin{tikzcd}
        L^2(G)
        \arrow{r}{\cF_{\cal I}}
        \arrow{d}{\rho^{\cal I}_{\text{pos}}(\hat A)}
        &
        \hat P_{\cal I}\cdot L^2_\star(\mathfrak{g}^*)
        \arrow{d}{\rho^{\cal I}_{\text{mom}}(\hat A)}
        \\
        L^2(G)
        \arrow{r}{\cF_{\cal I}}
        &
        \hat P_{\cal I}\cdot L^2_\star(\mathfrak{g}^*)
     \end{tikzcd}
     \label{eq: commutative diagram of F}
\end{equation}
Here $\rho_{\text{pos}}$ and $\rho_{\text{mom}}$ are the `large' position and momentum representations of the operator algebra $\cal A$, respectively defined in sections \ref{ssec: position rep 2} and \ref{ssec: momentum rep 1}. Their analogues with a superscript $\cal I$ are the `reduced' or projected representations of sections \ref{ssec: position rep 1} and \ref{ssec: momentum rep 2}, respectively. The same commutative diagram thus holds in both projected and unprojected cases. We therefore first build the Fourier \i{transform} $\cF$ by working in the larger Hilbert spaces $L^2(\mathfrak{g})$ and $L^2_\star(\mathfrak{g}^*)$, then project down to the smaller $\mathcal{I}$-invariant subspaces $L^2(G)$ and $\hat P_{\cal I}\cdot L^2_\star(\mathfrak{g}^*)$ to obtain Fourier \i{coefficients} and Fourier \i{series}. Most properties below will be stated for Fourier coefficients (with infinite $|\cal I|$), as their analogues for Fourier transforms are straightforward upon taking $\mathcal{I}=\{0\}$ everywhere.

\subsection{Noncommutative Fourier transform}
\label{sseFour}

Let us first show the existence of the intertwiner $\cF$ mapping $L^2(\mathfrak{g})$ on $L^2_\star(\mathfrak{g}^*)$, without reference to quotients by the identity subset \eqref{ix} due to the possible presence of compact subgroups of $G$. To that end, write $\cF$ as an integration kernel: for any wave function $\psi\in L^2(\mathfrak{g})$,
\begin{align}
\label{e56}
    \cF[\psi](p)
    =
    \int_\mathfrak{g} J(X)\dd^n X\; E(X,p)\psi(X)\in L^2_\star(\mathfrak{g}^*).
\end{align}
The integration kernel $E(X,p)$ is unknown at this stage, and needs to be found. It must be such that $\cF\circ \rho_{\text{pos}}(\hat A) = \rho_{\text{mom}}(\hat A)\circ\cF$ for any operator $\hat A\in\cal A$, as in the commutative diagram \eqref{eq: commutative diagram of F}. This must hold, in particular, for position and momentum operators. Hence, for any $\psi\in L^2(\mathfrak{g})$, one must have
\begin{align}
\int  J(X)\dd^n X\; i\frac{\partial E(X,p)}{\partial p_j} \psi(g)&= \int  J(X)\dd^n X\; E(X,p) X^j\psi(X),
     \label{eq: intertwiner position}\\
      \int  J(X)\dd^n X\; p_j\star E(X,p)\psi(X)&= \int  J(X)\dd^n X\; E(X,p)\left(-i \mathcal{L}_j\psi\right)(X),
      \label{eq: intertwiner momentum}    
\end{align}
where we used the momentum representation \eqref{eq: momentum op in momentum space}--\eqref{eq: zeta momentum rep} on the left-hand side, and the position representation \eqref{e20} on the right-hand side. Now integrate by parts on the right-hand side of \eqref{eq: intertwiner momentum}, keeping in mind that the Haar measure is left-invariant. Using the fact that \eqref{eq: intertwiner position}--\eqref{eq: intertwiner momentum} must hold for any $\psi$, one concludes that
\begin{equation}
i\frac{\partial E(X,p)}{\partial p_j}=  X^j E(X,p),
\qquad
p_j\star E(X,p)= i  \mathcal{L}_j E(X,p).
\label{eq: ode g of exp}
\end{equation}
Here the first equation is readily integrated into
\begin{align}
     E(X,p)= f(X)e^{-i\langle p,X\rangle},
     \label{eq: integrated ode p up to cst}
\end{align}
which fixes the momentum-dependence of the kernel in \eqref{e56}. To determine $f(X)$, integrate the second equation in \eqref{eq: ode g of exp} along right-invariant vector fields to get
\begin{align}
      E(B(-Y,X),p) = e^{-Y^i\cdot \mathcal{L}_i}E(X,p) 
      =
      e^{-i\langle p,Y\rangle}\star E(X,p).
      \label{eq: integrated ode g}
\end{align}
Plugging this back into \eqref{eq: integrated ode p up to cst} yields $f(B(-Y,X))=f(X)$ for all $Y$, so $f(X)$ is in fact constant. We will set this constant to 1 without loss of generality, so the kernel \eqref{eq: integrated ode p up to cst} of the noncommutative Fourier transform is a \i{plane wave}
\begin{align}
\label{e63}
     E(X,p)=e^{-i\langle p,X\rangle}.
\end{align}
As a result, the \i{noncommutative Fourier transform} \eqref{e56} coincides with the standard one on $\mathbb{R}^n$, save for an additional Jacobian \eqref{e19} due to the Haar measure:
\begin{align}
     \cF[\psi](p)
     :=
     \int_\mathfrak{g} J(X)\dd^n X \;e^{-i\langle p, X\rangle}\psi(X).
     \label{eq: def ncft}
\end{align}
This will be our definition of $\cF$. We stress that the exponential appearing here is the usual one, as opposed to a noncommutative $\star$-exponential as in Wigner-Weyl calculus \cite{Zachos}. This simplification is due to the fact that exponential functions are mapped to exponential operators under the symmetric ordering \eqref{expo} of the $p_i$s.

\paragraph{Properties of plane waves.} Let us list a few key properties of the plane waves \eqref{e63}. They are mostly self-evident, but our goal will be to reproduce them in the more complicated $\mathcal{I}$-invariant case. First, one has
\begin{equation}
E(X,0)=1,
\qquad
E(0,p)=1,
\qquad
E(-X,p) =\overline{E(X,p)} = E(X,-p),
\label{s15}
\end{equation}
which straightforwardly stem from \eqref{e63}. One more property involves star products, namely
\begin{align}
E(X,p)\star_p E(Y,p+q) &= E(B(X,Y),p)\,E(Y,q),
\label{e65}
\end{align}
where $B(X,Y)$ is the Baker-Campbell-Hausdorff expansion \eqref{baker}. This stems from the star product of plane waves implied by eq.~\eqref{eq: BCH} and the choice of ordering \eqref{expo}. It states that plane waves are well-behaved under both `multiplication' in $\mathfrak{g}$ and momentum addition. A special case of \eqref{e65} that will repeatedly be useful is $E(X,p)\star E(Y,p)= E(B(X,Y),p)$.

One last property will be essential in section \ref{ssec: prop of ncft}. Namely, the momentum integral of eq.~\eqref{e63} yields
\begin{equation}
\int_{\mathfrak{g}^*} \frac{\dd^n p}{(2\pi)^n}\; e^{-i\langle p, X\rangle}
=
\delta^n(X)
=
\delta_\mathfrak{g}(X),
 \label{eq: FT inv 1 is delta}
\end{equation}
where $\delta^n$ is the Dirac distribution for the Lebesgue measure on $\mathfrak{g}$, and $\delta_\mathfrak{g}$ is the Dirac distribution for the left-invariant Haar measure on the Baker-Campbell-Hausdorff group $\mathfrak{g}$. The second equality stems from the property $\int_\mathfrak{g} J(X)\dd^n X \psi(X)\delta^d(X)=J(0)\psi(0)=\psi(0)$, valid for any test function $\psi$ thanks to the fact that $J(0)=1$ owing to eq.~\eqref{e19}. A closely related statement is that the integration kernel
\begin{equation}
      \delta_\star(p,q)
      :=
      \int_\mathfrak{g} J(X)\dd^n X\;E(X,p)E(X,-q)
      \label{eq: big momentum delta}
\end{equation}
is such that the following holds for any test function $\Phi$ in momentum space:
\begin{equation}
\label{s19}
      \Phi(p)=\int_{\mathfrak{g}^*}\frac{\dd^n q}{(2\pi)^n}\; \delta_\star(p,q)\star_{q}\Phi(q).
\end{equation}
Indeed, it suffices to show that \eqref{s19} holds for plane waves $\Phi(p)=e^{-i\langle p,X\rangle}$, since they span $L^2_{\star}(\mathfrak{g}^*)$; this is the case thanks to \eqref{e65} and the property \eqref{eq: FT inv 1 is delta}.

\subsection{Noncommutative Fourier coefficients}
\label{ssec: noncommutative Fourier coefficients}
We now present the $\cal I$-invariant version of section \ref{sseFour}. A consequence of $\cal I$-invariance is that the Fourier \i{transform} becomes a map $\cF_{\cal I}$ that sends a function $\psi$ on its Fourier \i{coefficients} in momentum space. These `coefficients' turn out to be functions on specific submanifolds of momentum space $\mathfrak{g}^*$; they reduce to actual coefficients (labelled by discrete values of $p$) only when the group $G$ is both compact and Abelian, as in the standard case of Fourier series with $G=\text{U}(1)$ (see section \ref{sec: examples}).

\paragraph{$\cal I\!\!\!\!\cal I$-invariant plane waves.} To obtain non\-com\-mu\-ta\-tive Fou\-rier coef\-fi\-cients from the Fou\-rier trans\-form \eqref{eq: def ncft}, one imposes $\mathcal{I}$-invariance of $\psi$ so that $\cF[\psi]$ lies in the $\cal I$-invariant subspace $\hat P_{\cal I}\cdot L^2_\star(\mathfrak{g}^*)$, with $\hat P_{\cal I}$ the projector \eqref{proji}. This motivates the definition of \i{$\cal I$-invariant plane waves}
    \begin{align}
       E_\mathcal{I}(X,p) &
       :=
       \rho_{\text{pos}}(\hat P_\mathcal{I})E(X,p)
       =
       \rho_{\text{mom}}(\hat P_\mathcal{I})E(X,p),
       \label{s16}
\end{align} 
where $\rho_{\text{pos}}$ and $\rho_{\text{mom}}$ are the position and momentum representations respectively defined in \eqref{e20} and \eqref{eq: momentum op in momentum space}--\eqref{eq: zeta momentum rep}. Since both representations of the projector may be used, we will omit the representation map below and simply let $\hat P_{\cal I}:=\rho(\hat P_{\cal I})$. Using the explicit series \eqref{proji}, one can also write
\begin{align}
       E_\mathcal{I}(X,p) &
       =
       \frac{1}{|{\cal I}|}\sum_{Z\in \operatorname{Logs}(e)}e^{-i\langle p,B(Z,X)\rangle}
=
\frac{1}{\big|\operatorname{Logs}(\exp(X))\big|}\sum_{Y\in \operatorname{Logs}(\exp(X))}e^{-i\langle p,Y\rangle}
       \label{s16bis}
\end{align} 
where the set $\operatorname{Logs}(g)$ was defined above \eqref{ix}. The first rewriting is most convenient for abstract proofs, while the second is best for concrete computations.

One may be worried at this point by the appearance of infinite factors such as $|\mathcal{I}|$ or $|\operatorname{Logs}(\exp(X))|$ in eqs.~\eqref{s16bis}. In fact, these factors are nothing out of the ordinary, as they already appear when relating standard Fourier transforms and Fourier series. The factor $|\operatorname{Logs}(\exp(X))|$, in particular, is much smaller than the $|\mathcal{I}|$ factor for almost every $X$. Indeed, as discussed around \eqref{e84}, there are many more logarithms \eqref{eq: sum over I unpacked} of the identity than of a generic element of $G$. The factor $|\mathbb Z^r|$ that arises from \eqref{e84} is much more manageable than possibly divergent integrals along adjoint orbits, such as those occurring in $|\mathcal I|$ owing to eqs.~\eqref{eq: sum over I unpacked}. Hence, for almost every $X$, the $\cal I$-invariant plane wave \eqref{s16bis} simplifies to
\begin{align}
    E_\mathcal{I}(X,p)
    =
    \frac{e^{-i\langle p,X\rangle}}{|\mathbb{Z}|^r}\sum_{n^1,...,n^r\in \mathbb{Z}}e^{-i\langle p,2\pi n^i a_i(X)\rangle}
    =
    e^{-i\langle p, X\rangle}\sum_{k_1,...,k_r\in \mathbb{Z}} \prod_{i=1}^r\frac{\delta(\langle p, a_i(X)\rangle-k_i)}{|\mathbb Z|},
    \label{e85}
\end{align}
where we used the standard Poisson summation formula (see again section \ref{sec: examples}). It becomes clear in this way that the infinite $|\mathbb{Z}|$ factor regularizes a $\delta(0)$. Thus, for almost every $X$, one has
\begin{align}
    E_\mathcal{I}(X,p) = 
    \begin{cases}
        e^{-i\langle p, X\rangle} &\text{if }\langle p, a_i(X)\rangle\in \mathbb{Z}\text{ for all $i\in \{1,\dots r\}$},
        \\
        0 &\text{otherwise,}
    \end{cases}
    \label{e86}
\end{align}
which is nothing but a noncommutative version of a discrete Fourier mode, compatible with the Lie algebra $\mathfrak{g}$.

\paragraph{Fourier coefficients; comparison with \cite{Oriti:2011ac,Raasakka:2011np,Guedes:2013vi}.} Recalling the factor $1/\sqrt{|\mathbb Z|^r}$ in the correspondence \eqref{e5354} between functions on $G$ and periodic functions on $\mathfrak{g}$, define the \i{noncommutative Fourier coefficients}
\begin{equation}
\label{e68}
\begin{split}
      \cF_{\cal I}[\psi](p)
       &:=
       \frac{1}{\sqrt{|\mathbb Z|^r}}\int_\mathfrak{g} J(X)\dd^n X\; E_\mathcal{I}(X,p)\psi(X)\\[.3em]
       &{\color{white}{:}}=
       \sqrt{|\mathbb Z|^r}
       \!\!\!\!\int\limits_{\substack{\text{principal}\\ \text{branch in $\mathfrak{g}$}}} \!\!\!\!J(X)\dd^n X\,E_\mathcal{I}(X,p)\psi(X)
\end{split}
\end{equation}
for any periodic wave functions $\psi\in\hat P_{\cal I}\cdot L^2(\mathfrak{g})=L^2(G)$, where the principal branch of the logarithm was defined above eq.~\eqref{eq: prop of pb coordinates}. This is just the $\cal I$-invariant projection of the Fourier transform \eqref{eq: def ncft}. Owing to eq.~\eqref{e85}, it vanishes for almost any momentum $p\in\mathfrak{g}^*$, which is the sense in which it generalizes to any group $G$ the standard Fourier coefficients that occur for $G=\text{U}(1)$. It can be made more explicit by identifying the integral over the principal branch with an integral over the group as in eq.~\eqref{e18}, so that
\begin{align}
      \cF_{\cal I}[\psi](p)
      =
\sqrt{|\mathbb Z|^r}\int_{G}\dd g \;E_\mathcal{I}(g,p)\psi(g).
       \label{eq: best def Fourier series}
\end{align}
Here we let $E_\mathcal{I}(g,p):=E_\mathcal{I}(\log(g),p)$ since the sum over logarithms in \eqref{s16bis} ensures that the value of $E_\mathcal{I}(\log(g),p)$ does not depend on the choice of logarithm. As it turns out, the expression \eqref{eq: best def Fourier series} of noncommutative Fourier coefficients will be the most convenient one below. This is despite the infinite $\sqrt{|\mathbb Z|^r}$ factor, which may seem problematic at first sight, but actually ensures that the noncommutative Fourier coefficients $\cF_{\cal I}[\psi]$ are square-integrable when $\psi$ is square-integrable. In terms of plane waves, one can multiply eq.~\eqref{e85} by $\sqrt{|\mathbb Z|^r}$ to get
\begin{equation}
    \sqrt{|\mathbb Z|^r} E_\mathcal{I}(X,p)
    =
    e^{-i\langle p, X\rangle}\sum_{k_1,...,k_r\in \mathbb{Z}} \prod_{i=1}^r\sqrt{\delta(\langle p, a_i(X)\rangle-k_i)}
    \label{squaredelta}
\end{equation}
where we formally write $\sqrt{\delta(k)}:=\delta(k)/\sqrt{\delta(0)}$ and use the prescription $\delta(0):=|\mathbb Z|$, which is well justified in the U(1) case (see section \ref{ssec: U(1) case}). The square root of the delta function guarantees that plane waves have finite norm, \i{i.e.}~are square-integrable: this is again a version of finite-volume regularization in quantum mechanics.

Note that the definition \eqref{e68} can be recast as
\begin{equation}
\cF_{\cal I}[\psi](p)
=
\sqrt{|{\mathbb Z}|^r}\,\hat P_{\cal I}\cdot
       \!\!\!\!\int\limits_{\substack{\text{principal}\\ \text{branch in $\mathfrak{g}$}}} \!\!\!\!J(X)\dd^n X\,E(X,p)\psi(X),
       \label{ourdef}
\end{equation}
where we used eq.~\eqref{s16} and $E(X,p)$ is the plane wave \eqref{e63}, without $\cal I$-invariance. This is similar to, but crucially different from, the definition used in earlier literature on the subject \cite{Freidel:2005bb,Joung:2008mr,Oriti:2011ac,Raasakka:2011np,Guedes:2013vi}. Indeed, Fourier series there (say in refs.~\cite{Oriti:2011ac,Raasakka:2011np,Guedes:2013vi} for definiteness) are defined by a version of eq.~\eqref{ourdef} \i{without} the projector $\hat P_{\cal I}$. The problem then is that scalar products and the action of operators are ill-defined on the image space, as discussed below eq.~\eqref{quop}. Including the projector $\hat P_{\cal I}$ as in \eqref{ourdef} circumvents the issue, with an outcome that has finite norm thanks to the $\sqrt{|\mathbb Z|^r}$ factor.

\paragraph{Properties of $\cal I\!\!\!\!\cal I$-invariant plane waves.} Let us list some of the key features of the $\cal I$-invariant plane waves \eqref{s16}, following a sequence similar to that of section \ref{sseFour}. First, consider the $\cal I$-invariant version of eqs.~\eqref{s15}, namely
\begin{equation}
E_\mathcal{I}(g ^{-1},p) = \overline{E_\mathcal{I}(g,p)} = E_\mathcal{I}(g,-p),
\qquad
E_\mathcal{I}(g,0)=1,
\qquad
E_\mathcal{I}(e,p)=\mathcal{Q}^{-1}(\hat P_\mathcal{I}).
\label{s17}
\end{equation}
These readily follow from the definition \eqref{s16}. Eq.~\eqref{e65} similarly becomes
\begin{align}
\label{ee73}
E_\mathcal{I}(g,p)\star_p E_\mathcal{I}(h,p+q) &=E_\mathcal{I}(gh,p)E_\mathcal{I}(h,q),
\end{align}
showing that $\cal I$-invariant exponentials represent the group law under $\star$-multiplication, and behave well under momentum addition. A key corollary is $E_\mathcal{I}(g,p)\star E_\mathcal{I}(h,p)= E_\mathcal{I}(gh,p)$.
\begin{center}
\begin{minipage}{.9\textwidth}
\footnotesize%
\i{Proof of \eqref{ee73}.} The definition \eqref{s16bis} of $\mathcal I$-invariant plane waves yields
\begin{align*}
& E_\mathcal{I}(g,p)\star_p E_\mathcal{I}(h,p+q)= \frac{1}{|\mathcal I|^2}\sum_{Z\in \operatorname{Logs}(e)}\sum_{W\in \operatorname{Logs}(e)}e^{-i\langle p, B(Z,X)\rangle}\star_p e^{-i\langle p, B(Y,W)\rangle}e^{-i\langle q, B(Y,W)\rangle}
\\
&=\frac{1}{|\mathcal I|^2}\sum_{Z\in \operatorname{Logs}(e)}e^{-i\langle p, Z\rangle} \star_p e^{-i\langle p, X\rangle}\star_p e^{-i\langle p, Y\rangle}\star_p\sum_{W\in \operatorname{Logs}(e)}e^{-i\langle p,W\rangle}e^{-i\langle q, B(Y,W)\rangle}
\end{align*}
where $g=\exp(X)$, $h=\exp(Y)$ and we used eq.~\eqref{e65} to split into four a product of two plane waves. The product $e^{-i\langle p, X\rangle}\star_p e^{-i\langle p, Y\rangle}$ can be recast as $e^{-i\langle p, B(X,Y)\rangle}$ owing to eq.~\eqref{e26}. The latter can again be used to note that the sum over $Z$ involves $e^{-i\langle p, B(Z,B(X,Y))\rangle}$, so
\begin{equation*}
E_\mathcal{I}(g,p)\star_p E_\mathcal{I}(h,p+q) 
=
E_{\mathcal I}(gh,p)\star_p\frac{1}{|\mathcal I|}\sum_{W\in \operatorname{Logs}(e)}e^{-i\langle p,W\rangle}e^{-i\langle q, B(Y,W)\rangle}.
\end{equation*}
Finally using the fact that $E_{\mathcal I}(gh,p)$ is $\cal I$-invariant by definition, one has $E_{\mathcal I}(gh,p)\star e^{-i\langle p,W\rangle}=E_{\mathcal I}(gh,p)$ provided $W$ is a logarithm of the identity (which it is by assumption). The remaining sum over $W$ reduces to an $\cal I$-invariant plane wave \eqref{s16bis}, which gives eq.~\eqref{ee73}.
\hfill$\blacksquare$
\end{minipage}
\end{center}
Finally, one last property is an $\cal I$-invariant analogue of eq.~\eqref{eq: FT inv 1 is delta}. Namely, the momentum integral of $\mathcal{I}$-invariant plane waves satisfies
\begin{equation}
     |\mathbb Z|^r\int_{\mathfrak{g}^*}\frac{\dd^n p}{(2\pi)^n}\; E_\mathcal{I}(g,p) = \int_{\mathfrak{g}^*}\frac{\dd^n p}{(2\pi)^n}\; E(\log(g),p) = \delta(g),
     \label{eq: inv fourier series of 1 is delta}
\end{equation}
where $\delta(g)$ is the delta distribution for the Haar measure on $L^2(G)$. To be more explicit, for any test function $\psi$ on $G$, one has
\begin{align}
\label{pseh}
      \psi(e)
      =
      \int_G\dd g
      \int_{\mathfrak{g}^*}\frac{\dd^np}{(2\pi)^n}
      \;E(\log (g),p)\psi(X)
      =
      |\mathbb Z|^r
      \int_G\dd g
      \int_{\mathfrak{g}^*}\frac{\dd^np}{(2\pi)^n}
      \;E_\mathcal{I}(g,p)\psi(X),
\end{align}
where $\psi(X)$ is $\psi(g)$ written in the principal branch coordinates defined around eq.~\eqref{eq: prop of pb coordinates}.
\begin{center}
\begin{minipage}{.9\textwidth}
{\footnotesize%
\i{Proof of \eqref{eq: inv fourier series of 1 is delta}.} The second equality in \eqref{eq: inv fourier series of 1 is delta}, which only involves regular plane waves, readily stems from the identity \eqref{eq: FT inv 1 is delta}. The first equality, by contrast, involves $\mathcal{I}$-invariant plane waves. Using again the identity \eqref{eq: FT inv 1 is delta}, one finds $\int_{\mathfrak{g}^*} \frac{\dd^n p}{(2\pi)^n}\;E_\mathcal{I}(X,p)=\frac{1}{|{\cal I}|}\sum_{Y\in \operatorname{Logs}(e)} \delta_\mathfrak{g}(B(Y,X))$ in terms of the Dirac distribution for the left Haar measure on the Baker-Campbell-Hausdorff group $\mathfrak{g}$. It follows, upon seeing any test function $\psi\in L^2(G)$ as an $\mathcal{I}$-periodic function on $\mathfrak{g}$, that
\begin{align*}
    \frac{|\mathbb Z|^r}{|{\cal I}|}\sum_{Y\in {\cal I}}\!\!\int\limits_{\substack{\text{principal}\\ \text{branch in $\mathfrak{g}$}}} \!\!\!\!\! J(X)\dd^nX \; \psi(X) \delta_\mathfrak{g}(B(Y,X)) 
    &= 
      \frac{1}{|{\cal I}|}\sum_{Y\in {\cal I}}\int_{\mathfrak{g}} J(X)\dd^nX \; \psi(X) \delta_\mathfrak{g}(B(Y,X)).
\end{align*}
Changing the integration variable on the right-hand side from $X$ to $B(Y,X)$, and integrating over the delta function, makes this equal to $\frac{1}{|{\cal I}|}\sum_{Y\in {\cal I}} \psi(-Y)=\psi(0)$, proving eq.~\eqref{pseh}.\hfill$\blacksquare$}
\end{minipage}
\end{center}

\subsection{Properties of noncommutative Fourier series}
\label{ssec: prop of ncft}

In the following, we always work with noncommutative Fourier coefficients \eqref{eq: best def Fourier series} as opposed to Fourier transforms \eqref{eq: def ncft}. The latter can indeed be recovered from the former upon replacing the group $G$ by its Lie algebra $\mathfrak{g}$ endowed with the Baker-Campbell-Hausdorff composition law \eqref{xy}, whereupon the $\cal I$-invariant exponentials \eqref{s16} reduce to plain exponentials \eqref{e63}, in which case ${\cal I}=\{0\}$, $|{\cal I}|=1$ and $r=0$, $|\mathbb Z|^r=1$.

The plan is to first derive basic properties of noncommutative Fourier coefficients, then introduce the inverse Fourier transform to define Fourier series, showing, as desired, that $\cF_{\cal I}$ and its inverse provide an isometry between the Hilbert spaces $L^2(G)$ and $\hat P_{\cal I}\cdot L^2_{\star}(\mathfrak{g}^*)$. These statements are well-known for commutative Fourier series, but their proof in the noncommutative case requires more work. Along similar lines, we show that standard properties involving Fourier transforms of convolutions remain true, up to the use of star products instead of pointwise multiplication of functions. We end by deriving the group-theoretic version of the `lone star lemma' of Wigner-Weyl calculus \cite{Zachos}.

\paragraph{Properties of Fourier coefficients.} Here we list some immediate properties of the Fourier coefficients \eqref{eq: best def Fourier series}. Each is presented as a statement followed by a proof.

\noindent\i{Statement.} Noncommutative Fourier coefficients preserve scalar products in $L^2(G)$ and $\hat P_{\cal I}\cdot L^2_\star(\mathfrak{g}^*)$. Explicitly, if $\phi,\psi$ are wave functions in $L^2(G)$, then
\begin{equation}
\label{e76}
\langle\phi|\psi\rangle_G
=
\big\langle \cF_{\cal I}[\phi]\big| \cF_{\cal I}[\psi]\big\rangle_{\mathfrak{g}^*}
\end{equation}
where the pairings on the left- and right-hand sides were respectively defined in eqs.~\eqref{e18} and \eqref{eq: scalar product momentum space}. As a corollary, the pairing \eqref{eq: scalar product momentum space} is positive-definite on the image of the Fourier transform, \i{i.e.}~it is a genuine scalar product.
\begin{center}
\begin{minipage}{.9\textwidth}
{\footnotesize%
\i{Proof.} Starting from the right-hand side of \eqref{e76}, use the Fourier coefficients \eqref{eq: best def Fourier series} and the scalar product \eqref{eq: scalar product momentum space} to get
\begin{equation}
\big\langle  \cF_{\cal I}[\phi]\big| \cF_{\cal I}[\psi]\big\rangle_{\mathfrak{g}^*}
= 
|\mathbb Z^r|\int \frac{\dd^np\,\dd g\,\dd h}{(2\pi)^n}\;\bar\phi(g)\psi(h)E_\mathcal{I}(g ^{-1},p)\star E_\mathcal{I}(h,p).
\nonumber
\end{equation}
Now use the fact that exponentials represent the group law as in \eqref{ee73}, along with the momentum integral \eqref{eq: inv fourier series of 1 is delta}, to write $\langle\cF_{\cal I}[\phi]|\cF_{\cal I}[\psi] \rangle_{\mathfrak{g}^*}=\int \dd g\,\dd h \;\bar\phi(g)\psi(h)\delta(g ^{-1}h)=\int \dd g \;\bar\phi(g)\psi(g)$. This is nothing but the scalar product \eqref{e18} in $L^2(G)$. \hfill$\blacksquare$}
\end{minipage}
\end{center}
\noindent\i{Statement.} Noncommutative Fourier coefficients are well-behaved under left translations: for any wave function $\psi\in L^2(G)$ and any $g\in G$, one has
\begin{align}
    \cF_{\cal I}[\psi(g\;\cdot\,)](p)
    =
    E_\mathcal{I}(g ^{-1},p)\star \cF_{\cal I}[\psi](p).
\end{align}
\begin{center}
\begin{minipage}{.9\textwidth}
{\footnotesize%
\i{Proof.} The definition \eqref{eq: best def Fourier series} of Fourier coefficients yields 
$
\cF_{\cal I}[\psi(g\,\cdot\,)](p)
=
\sqrt{|\mathbb Z|^r}\int_G\dd h\,E_{\cal I}(h,p)\psi(gh)
=
\sqrt{|\mathbb Z|^r}\int_G\dd h\,E_{\cal I}(g^{-1}h,p)\psi(h)
=
E_{\cal I}(g^{-1},p)\star \cF_{\cal I}[\psi](p)
$, where we used \eqref{ee73} and the left-invariance of the Haar measure on $G$. \hfill$\blacksquare$}
\end{minipage}
\end{center}
\noindent\i{Statement.} The noncommutative Fourier coefficients of the Haar delta distribution $g\mapsto\delta(g)$ are given by an $\cal I$-invariant exponential at the identity:
\begin{align}
      \cF_{\cal I}\!\bigg[\frac{\delta}{\sqrt{|\mathbb Z|^r}}\bigg](p)
      =
      E_{\cal I}(e,p),
      \qquad\text{that is,}\qquad
      \cF_{\cal I}\!\bigg[\frac{\delta}{\sqrt{|\mathbb Z|^r}}\bigg]\star \Phi(p) 
      =
      \Phi(p)
       \label{eq: fourier delta is 1}
\end{align}
     for any $\cal I$-invariant wave function $\Phi$ in momentum space. Thus, the Fourier transform of the Dirac distribution at the identity is just $1$, similar to the usual Dirac distribution on $\mathbb{R}^n$.
\begin{center}
\begin{minipage}{.9\textwidth}
{\footnotesize%
\i{Proof.} This one is trivial: apply the definition \eqref{eq: best def Fourier series} to $\psi(g)=\delta(g)$. \hfill$\blacksquare$}
\end{minipage}
\end{center}

\paragraph{Inverse Fourier transform.} Define the \i{adjoint} of the Fourier coefficients $\cF_{\cal I}$ to be the \i{Fourier series} $\cF_{\mathcal{I}}^{\dagger}:\hat P_{\cal I}\cdot L_\star^2(\mathfrak{g}^*)\rightarrow L^2(G)$ given by
\begin{align}
      \cF_{\mathcal{I}}^{\dagger}[\Phi](g)
      :=
      \sqrt{|\mathbb Z|^r}\int_{\mathfrak{g}^*} 
      \frac{\dd^n p}{(2\pi)^n}\;\overline{E_\mathcal{I}(g,p)}\star\Phi(p)
      =\sqrt{|\mathbb Z|^r}\int_{\mathfrak{g}^*} \frac{\dd^n p}{(2\pi)^n}
      \;\overline{E(g,p)}\star\Phi(p),
      \label{defad}
\end{align}
where the second equality holds because $\Phi$ is $\mathcal{I}$-invariant by assumption. Let us now prove that $ \cF_{\mathcal{I}}^{\dagger}$ is the inverse of $ \cF_{\mathcal{I}}$.

\noindent\i{Statement.}\vspace{-2.1em}
\begin{align}
\label{e80}
      \cF_{\mathcal{I}}^{\dagger}\circ \cF_{\cal I}
      =
      \mathbb{I}_{L^2(G)}.
\end{align}
\begin{center}
\begin{minipage}{.9\textwidth}
{\footnotesize%
\i{Proof.} Pick any wave function $\psi\in L^2(G)$ and use eqs.~\eqref{s17}--\eqref{ee73} together with the plane wave representation of the delta function \eqref{eq: inv fourier series of 1 is delta} to find
\begin{align*}
~~~~~~~~~~\cF_{\mathcal{I}}^{\dagger}\big[\cF_{\cal I}[\psi]\big](g)
&=
|\mathbb Z|^r\int \frac{\dd^n p\, \dd h}{(2\pi)^n}\;\overline{E_\mathcal{I}(g,p)}\star E_{\mathcal I}(h,p) \psi(h)
\\
&= |\mathbb Z|^r\int \frac{\dd^n p\, \dd h}{(2\pi)^n}\;E_\mathcal{I}(g^{-1}h,p) \psi(h) = \int \dd h \;\delta(g^{-1}h)\psi(h) = \psi(g).~~~~~~~~~~~\hspace{.4em}\blacksquare
\end{align*}}
\end{minipage}
\end{center}
\noindent\i{Statement.} The Dirac distribution in momentum space can be represented as 
\begin{align}
\delta_\star^\mathcal{I}(p,q)= |\mathbb Z|^r\int_G \dd g\; E_\mathcal{I}(g,p)E_\mathcal{I}(g,-q),
\label{small momentum delta}
\end{align}
meaning that for $\Phi$ an $\mathcal{I}$-invariant wave function, $\int_{\mathfrak{g}^*}\frac{\dd^n q}{(2\pi)^n} \delta_\star^\mathcal{I}(p,q)\star_q \Phi(q)= \Phi(p)$. As a corollary, one has
\begin{align}
\label{s17q}
     \cF_{\cal I}\circ \cF_{\mathcal{I}}^{\dagger} = \mathbb{I}_{\hat P_{\cal I}\cdot L^2_\star(\mathfrak{g}^*)}.
\end{align}
\begin{center}
\begin{minipage}{.9\textwidth}
{\footnotesize%
\i{Proof.} Owing to the definitions \eqref{eq: best def Fourier series} and \eqref{defad} of $\cF_{\cal I}$ and its adjoint, the composition $\cF_{\cal I}\circ \cF_{\mathcal{I}}^{\dagger}$ acts on any $\cal I$-invariant wave function $\Phi$ on $\mathfrak{g}^*$ through the kernel \eqref{small momentum delta}:
\begin{align*}
\cF_{\cal I}\big[\cF_{\mathcal{I}}^{\dagger}[\Phi]\big](p) 
&=
|\mathbb Z|^r\int_{\mathfrak{g}^*} \frac{\dd^nq}{(2\pi )^n} \left(\int_G \dd g\;E_\mathcal{I}(g,p)E_\mathcal{I}(g,-q)\right)\star_q\Phi(q)
=
\int_{\mathfrak{g}^*} \frac{\dd^nq}{(2\pi )^n} \delta_\star^\mathcal{I}(p,q)\star_q\Phi(q).
\end{align*}
Our goal is therefore to show that the kernel \eqref{small momentum delta} is a delta function when acting on $\cal I$-invariant test functions. That this is the case follows from the simpler property \eqref{s19}, valid for functions on the whole of $L^2_\star(\mathfrak{g}^*)$. Seeing functions on the group as `periodic' functions on the Lie algebra, one can write $|\mathbb Z|^r\int_G \dd g = \int_\mathfrak{g} J(X)\dd^n X$, leading to
\begin{align*}
\cF_{\cal I}\big[\cF_{\mathcal{I}}^{\dagger}[\Phi]\big](p)
=
\int_{\mathfrak{g}^*} \frac{\dd^nq}{(2\pi )^n} \left(\int_\mathfrak{g} J(X)\dd^n X\;E_\mathcal{I}(X,p)E_\mathcal{I}(X,-q)\right)\star_q\Phi(q).
\end{align*}
Finally using the fact that $\Phi$ is $\mathcal{I}$-invariant and the definition \eqref{s16} of $\mathcal{I}$-invariant plane waves, one finds
\begin{align*}
    \cF_{\cal I}\big[\cF_{\mathcal{I}}^{\dagger}[\Phi]\big](p)
    =
    \hat P_\mathcal{I}\int_{\mathfrak{g}^*} \frac{\dd ^dq}{(2\pi )^d} \left(\int_\mathfrak{g} J(X)\dd^n X\;E(X,p)E(X,-q)\right)\star \Phi(q) \overset{\eqref{eq: big momentum delta}}{=}\hat P_\mathcal{I}\Phi(p)= \Phi(p).
\end{align*}This proves the desired properties \eqref{small momentum delta}--\eqref{s17q}. \hfill$\blacksquare$}
\end{minipage}
\end{center}
\noindent\i{Statement.} Noncommutative Fourier series are well-behaved under momentum addition: for any $\cal I$-invariant wave function $\Phi\in\hat P_{\cal I}\cdot L^2_{\star}(\mathfrak{g}^*)$ and for any momentum $q\in\mathfrak{g}^*$, one has
\begin{equation}
\cF_{\mathcal{I}}^{\dagger}[\Phi(\,\cdot+q)](g)
=
E_\mathcal{I}(g,q)\cF_{\mathcal{I}}^{\dagger}[\Phi](g).
\end{equation}
\begin{center}
\begin{minipage}{.9\textwidth}
{\footnotesize%
\i{Proof.}
The definition \eqref{defad} of Fourier series $\cF_{\mathcal{I}}^{\dagger}[\Phi(\,\cdot+q)](g)
=
\sqrt{|Z|^r}\int \frac{\dd^n p}{(2\pi)^n}\; \overline{E_\mathcal{I}(g,p)}\star_p \Phi(p+q)$. By the first equation in \eqref{s17}, write $\overline{E_\mathcal{I}(g,p)}=E_\mathcal{I}(g ^{-1},p)$. Further using the fact that the Fourier coefficients \eqref{eq: best def Fourier series} are surjective in $\hat P_{\cal I}\cdot L^2_{\star}(\mathfrak{g}^*)$, write $\cF_{\mathcal{I}}^{\dagger}[\Phi(\cdot+q)](g)=|\mathbb Z^r|\int \frac{\dd h\,\dd^n p}{(2\pi)^n}\; E_\mathcal{I}(g ^{-1},p)\star_p E_\mathcal{I}(h,p+q)\cF_{\mathcal{I}}^{\dagger}[\Phi](h)$. Finally use eq.~\eqref{ee73} to get
\begin{equation}
~~~~~~~\cF_{\mathcal{I}}^{\dagger}[\Phi(\,\cdot+q)](g)
=
|\mathbb Z|^r\int \frac{\dd h\,\dd^n p}{(2\pi)^n}\; E_\mathcal{I}(g ^{-1}h,p) E_\mathcal{I}(h,q)\cF_{\mathcal{I}}^{\dagger}[\Phi](h)
=
E_\mathcal{I}(g,q)\cF_{\mathcal{I}}^{\dagger}[\Phi](g).~~~~~\hspace{1em}~~\blacksquare
\nonumber
\end{equation}}
\end{minipage}
\end{center}
\noindent\i{Statement.} The Fourier series of the Dirac distribution $p\mapsto \delta_\star^\mathcal{I}(p,q)$ is an ${\cal I}$-invariant exponential
\begin{equation}
 \cF_{\mathcal{I}}^{\dagger}\!\bigg[\frac{\delta_\star(\,\cdot\,,q)}{\sqrt{|\mathbb Z|^r}}\bigg](g)
 =
 E_\mathcal{I}(g,-q).
       \label{eq: inv fourier delta is one}
\end{equation}
\begin{center}
\begin{minipage}{.9\textwidth}
{\footnotesize%
\i{Proof.} First note that $\overline{\delta_\star^\mathcal{I}(p,q)}=\delta_\star^\mathcal{I}(q,p)$. Using \eqref{eq: prop star under conjugation}, this implies that 
$\cF_{\mathcal{I}}^{\dagger}\!\Big[\frac{\delta_\star(\,\cdot\,,q)}{\sqrt{|{\cal I}|}}\Big](g)
=\int\frac{\dd^n p}{(2\pi)^n}\;\overline{E_\mathcal{I}(g,p)}\star \delta_\star(p,q)
=\overline{\int\frac{\dd^n p}{(2\pi)^n}\;\delta_\star(q,p)\star E_\mathcal{I}(g,p)}=E_\mathcal{I}(g,-q)$.\hfill$\blacksquare$}
\end{minipage}
\end{center}

\paragraph{Products and Fourier series.} Similarly to standard Fourier transforms, their noncommutative cousins convert convolutions of functions into products, and vice-versa. The only subtlety is the appearance of star products. Relatedly, the `lone star lemma' of Wigner-Weyl calculus still holds in a slightly modified form. We now list these properties:

\noindent\i{Statement.} Given functions $\phi,\psi$ on $G$ that decay sufficiently fast at infinity if $G$ is noncompact, define their \i{left convolution} by\footnote{The convolution product \eqref{tt23} generally fails to converge when $\phi,\psi\in L^2(G)$.}
\begin{equation}
\label{tt23}
(\phi * \psi)(g):=\int_G \dd h\; \phi(h^{-1}g)\psi(h).
\end{equation}
Then, the Fourier coefficients of the convolution satisfy
\begin{align}
\label{e80bis}
\cF_{\cal I}[\phi*\psi]
=\frac{1}{\sqrt{|\mathbb Z|^r}}
\cF_{\cal I}[\phi]\star\cF_{\cal I}[\psi]
\end{align}
where $\star$ is the momentum-space star product defined in \eqref{eq: def star product}.
\begin{center}
\begin{minipage}{.9\textwidth}
{\footnotesize%
\i{Proof.} Applying the definition \eqref{eq: best def Fourier series} of Fourier series to the convolution \eqref{tt23} yields $\cF_{\cal I}[\phi * \psi](p)=\sqrt{|\mathbb Z|^r}\int \dd g\dd h\; E_\mathcal{I}(g,p)\phi(h ^{-1}g)\psi(h)= \sqrt{|\mathbb Z|^r}\int \dd g\dd h\; E_\mathcal{I}(h ^{-1}g,p)\star E_\mathcal{I}(h,p)\phi(h ^{-1}g)\psi(h)$, where we used eq.~\eqref{ee73} at $q=0$. Now exploit the invariance of the Haar measure to change the integration variable from $g$ to $gh^{-1}$, which gives eq.~\eqref{e80bis}.\hfill$\blacksquare$}
\end{minipage}
\end{center}
\noindent\i{Statement.} Define the convolution of momentum-space functions in $\hat P_\mathcal{I}\cdot L^2_{\star}(\mathfrak{g}^*)$ by 
\begin{align}
\label{e81}
(\Phi*\Psi)(p)
:=
\sqrt{|\mathbb Z|^r}\int \frac{\dd^n q}{(2\pi)^n}\;\Phi(p-q)\star\Psi(q),
\end{align}
where the factor $\sqrt{|\mathbb Z|^r}$ ensures that the result is square-integrable. Then, the Fourier series of the convolution satisfies
\begin{align}
\label{e82}
\cF_{\cal I}^{\dagger}[\Phi*\Psi]
=
\cF_{\cal I}^{\dagger}[\Phi]\cF^{\dagger}_{\cal I}[\Psi].
\end{align}
\begin{center}
\begin{minipage}{.9\textwidth}
{\footnotesize%
\i{Proof.} Applying the definition \eqref{defad} of Fourier series to the convolution \eqref{e81} yields $\cF_{\mathcal{I}}^{\dagger}[\Phi *\Psi](g)=
|\mathbb Z|^r\int \frac{\dd^np \,\dd^nq}{(2\pi)^{2n}}\;\overline{E_\mathcal{I}(g,p)}\star_p\Phi(p-q)\star_{q}\Psi(q)$. We simplify notation by letting $\phi:=\cF_{\mathcal{I}}^{\dagger}[\Phi]$ and similarly for $\Psi$. Then using eqs.~\eqref{s17} and the surjectivity of Fourier coefficients yields
\begin{equation}
\cF_{\mathcal{I}}^{\dagger}[\Phi *\Psi](g)
=
|\mathbb Z|^{2r}\int \frac{\dd h\,\dd h'\,\dd^n p\,\dd^nq}{(2\pi)^{2d}}\;E_\mathcal{I}(g ^{-1},p)\star_p E_\mathcal{I}(h,p-q)\star_{q}E_\mathcal{I}(h',q)\;\phi(h)\psi(h').
\nonumber
\end{equation}
Now use \eqref{ee73} and the Haar delta distribution \eqref{eq: inv fourier series of 1 is delta} to get eq.~\eqref{e82}.\hfill$\blacksquare$}
\end{minipage}
\end{center}
\noindent\i{Statement.} The lone star lemma holds: for any two wave functions $\Phi,\Psi$ that decay sufficiently fast at infinity in momentum space $\mathfrak{g}^*$, one has
\begin{align}
      \int_{\mathfrak{g}^*} \frac{\dd^n p}{(2\pi)^n}\;\bar \Phi\star \Psi(p) 
     =
     \int_{\mathfrak{g}^*} \frac{\dd^n p}{(2\pi)^n}\; \left(\frac{1}{J\!\left(-i\frac{\partial}{\partial p}\right)}\bar \Phi(p)\right) \Psi(p).
     \label{eq: lone star lemma}
\end{align}
Here $J$ is the Jacobian \eqref{e19}, and $f(-i\frac{\partial}{\partial p})$ denotes the differential operator obtained by Taylor-expanding the function $f(X)$ and replacing each monomial in the expansion by the corresponding differential operator. In particular, the inverse Fourier transform, given by \eqref{defad} with ${\cal I}=\{0\}$, can be recast as
\begin{align}
      \cF^{\dagger}[\Phi](g)
      =
      \frac{\sqrt{|\mathbb Z|^r}}{J(\log(g))}\int_{\mathfrak{g}^*}\frac{\dd^np}{(2\pi)^n} e^{i\langle p, \log(g)\rangle}\Phi(p)
      \label{eq: no star inv Fourier}
\end{align}
without any star product. This will be useful for Poisson summation in section \ref{sec: examples}.
\begin{center}
\begin{minipage}{.9\textwidth}
{\footnotesize%
\i{Proof.} The star product is bilinear, so it suffices to show the property for plane waves, for which eq.~\eqref{eq: FT inv 1 is delta}  yields $\int_{\mathfrak{g}^*} \frac{\dd^n p}{(2\pi)^n}\;\overline{E(X,p)}\star E(Y,p) 
= \int \frac{\dd^n p}{(2\pi)^n}\;E(B(-X,Y),p) 
=\delta_\mathfrak{g}(B(-X,Y))$. 
Now using the fact that the Haar delta function for the Baker-Campbell-Hausdorff group is related to its Lebesgue analogue by a Jacobian, one has
\begin{align*}
~~~~~~~~~~~~~~~~~~~~~~~~~~~\delta_\mathfrak{g}(B(-X,Y))
&=
\frac{\delta^d(X-Y)}{J(X)}
=
\frac{1}{J(X)}\int_{\mathfrak{g}^*} \frac{\dd^n p}{(2\pi)^n}\; e^{i\langle p,X-Y\rangle} \\
&=
\int_{\mathfrak{g}^*} \frac{\dd^n p}{(2\pi)^n}\; \left(\frac{1}{J\!\left(-i\frac{\partial}{\partial p}\right)}\overline{E(X,p)}\right) E(Y,p).~~~~~~~~~~~~~~~~\hspace{.3em}~~~~\blacksquare
\end{align*}}
\end{minipage}
\end{center}

\subsection{Quantum mechanics on Lie groups}
\label{ssequant}

The statements \eqref{e80} and \eqref{s17q} show that noncommutative Fourier coefficients and series intertwine the position and momentum representations of quantum mechanics on a Lie group $G$, respectively defined on the Hilbert spaces $L^2(G)$ and $\hat P_{\cal I}\cdot L^2_{\star}(\mathfrak{g}^*)$. In particular, one may now conclude that the pairing \eqref{eq: scalar product momentum space} is actually a positive-definite scalar product on $L^2_{\star}(\mathfrak{g}^*)$. Note that this is yet another difference between the present work and refs.~\cite{Oriti:2011ac,Raasakka:2011np,Guedes:2013vi}, where a different Fourier transform was shown to be injective but \i{not} surjective, so that no property such as \eqref{s17q} could be derived. (See in particular \cite[eq.~(4.16)]{Guedes:2013vi}.)

At this point, the stage is set for quantum mechanics on any Lie group. One may think of the Hilbert spaces $L^2(G)$ and $\hat P_{\cal I}\cdot L^2_{\star}(\mathfrak{g}^*)$ as being one and the same, which is typically done when dealing with quantum mechanics on $G=\mathbb{R}^n$. As is common practice, one may introduce a basis of position kets $|g\rangle$ and momentum kets $|p\rangle$ such that $\langle g|\psi\rangle=\psi(g)$ and $\langle p|\psi\rangle=\cF_{\cal I}[\psi](p)$ for any $\psi\in L^2(G)$, with $\cF_{\cal I}$ the noncommutative Fourier coefficients \eqref{eq: best def Fourier series}. The change of basis is given by $\bra{p}\ket{g} = \sqrt{|\mathbb Z|^r}E_\mathcal{I}(g,p)$. The identity operator can be written as
\begin{equation}
\mathbb{I} 
=
\int_G \dd g\,\ket g\bra g
=
\int_{\mathfrak{g}^*}\frac{\dd^np}{(2\pi)^n}\,\ket{p}\star \bra{p},
  \label{eq: id in momentum basis}
\end{equation}
respectively in position space and momentum space.

Such statements typically form the starting point needed for path integrals and Wigner functions, which will be treated separately \cite{Beauvillain2}. One can nevertheless anticipate the gist of the result as follows, focussing for definiteness on path integrals. Given a Hamiltonian operator $\hat H$ acting in $L^2(G)$, the first step is typically to write the corresponding propagator $\langle g_f|e^{-i\hat{H}T}|g_i\rangle$ as a path integral, with $g_i$ and $g_f$ any two `initial' and `final' points in $G$. This is achieved in the usual way, by splitting the time interval $[0,T]$ into a large number $N$ of subintervals, then letting $N$ go to infinity. Crucially, each subinterval is accompanied by an insertion of the identity \eqref{eq: id in momentum basis} and an appearance of the classical symbol of the Hamiltonian, $H(g,p):= \bra{g}\ket{p}\star\bra{p}\hat H\ket{g}$. Note that this requires the noncommutative plane waves and the star product defined above, as in \cite{Oriti:2011ac}. At the end of the day, the propagator becomes an expression of the form
\begin{equation}
\langle g_f|e^{-i\hat{H}t}|g_i\rangle
=
\int{\cal D}g\,{\cal D}p\,
\exp\left(i\int_0^T\text{d}t\Big[\langle p,g ^{-1}\partial_t g\rangle-H(g,p)\Big]\right)
\label{pint}
\end{equation}
where the integral is taken over all paths $(g(t),p(t))$ in phase space such that $g(0)=g_i$ and $g(T)=g_f$. The functional measure ${\cal D}g$ is an `infinite product' of Haar measures at each time $t$, and the measure ${\cal D}p$ is an analogous product of flat Lebesgue measures in $\mathfrak{g}^*\cong\mathbb{R}^n$. Some subtleties arise when $G$ has compact subgroups, so that the seemingly innocuous expression \eqref{pint} contains sums over images that can be treated by relating the `small' and `large' regular representations as in section \ref{ssec: position rep 2}. Relatedly, the mixed propagator $\langle p_f|e^{-i\hat{H}t}|g_i\rangle$ typically involves a nonabelian Poisson summation, treated here in section \ref{sec: examples}. These details will be covered in \cite{Beauvillain2}.

Starting from eq.~\eqref{pint}, it is straightforward to derive the classical limit of the propagator, or that of the canonical partition function $\int\text{d}g\,\langle g|e^{-\beta \hat{H}}|g\rangle$ at temperature $1/\beta$. This is because (after a field redefinition) the Hamiltonian action in the exponent in \eqref{pint} has a saddle point right at the equations of motion \eqref{3a}--\eqref{3b}. In this way, the path integral \eqref{pint} can be used to systematically study quantum corrections to Lie-Poisson dynamics---one of our original motivations for this project, and one that will be explored in greater detail in \cite{Beauvillain2}. A prime example of a system that can be described in this way is the quantum rigid body, whose Hilbert space is $L^2(\text{SO}(3))$ and whose energy spectrum cannot, in general, be written in closed form (see \textit{e.g.}~\cite{Allen}). The path integral approach provides an approximation scheme for, say, the partition function of the rigid body, circumventing the problem of finding its spectrum. Another system that can be described by a path integral such as \eqref{pint} is a (continuous and periodic) spin chain, whose configurations are orientations of an infinity of spins, each sitting at one point of a circle. These configurations, it turns out, can be seen as noncommutative `momenta' obtained by quantizing the symplectic reduction of a larger Hilbert space $L^2(L\text{SO}(3))$, where $L\text{SO}(3)$ is the loop group of $\text{SO}(3)$ \cite{Khesin}. Since quantization famously commutes with reduction \cite{Meinrenken1998}, one may hope to use the tools of the present paper to study geometric observables of spin chains in a way that admits a straightforward classical limit. (A more subtle questions is whether the classical limit commutes with both quantization and reduction; this is beyond our scope.)

Note that one could have attempted to derive a path integral in $L^2(G)$ by relying on the Peter-Weyl theorem, according to which the regular representation of $G$ decomposes into a sum of its irreducible unitary representations weighed by their multiplicity (see \textit{e.g.}~\cite{Williams} or \cite[sec.~3]{Sepanski}). The identity operator would then be written as in the first equality of \eqref{eq: id in momentum basis} in position space, but its expression in momentum space would involve instead a discrete sum over matrix elements of irreducible representations, which would then play the role of discrete Fourier modes. In fact, this is how we initially attempted to tackle the derivation of the path integral. The issue is that doing so makes it near-impossible to see any classical quantity appear, since `momenta' are discrete by definition. A perfect illustration is the particle on a circle, whose propagator (or its Euclidean counterpart, the partition function) can indeed be written in two equivalent ways, either as a sum over discrete Fourier modes, or as a genuine path integral. The former gives access to the low-temperature, quantum regime; the latter is more useful for the high-temperature, classical regime. The two are linked by Poisson summation, which converts a discrete sum into a sum of integrals with extra winding numbers. But, to our knowledge, no Poisson summation has been developed in general for arbitrary Lie groups, which is why we went through all this trouble here. In particular, section \ref{sec: examples} is devoted to a detailed discussion of this issue, with key applications in the context of path integrals \cite{Beauvillain2}.

\section{Applications of noncommutative Fourier series}
\label{sec: examples}

We now discuss two examples of Fourier series in detail, namely those appropriate to the groups U(1) and SU(2). Even though the former is Abelian, its compactness makes the construction nontrivial, illustrating the link between Fourier transforms and Fourier coefficients. The example of SU(2) furthermore exhibits all the subtleties of noncommutative Fourier series, leading to a new formula for the Poisson summation of functions in $\mathbb{R}^3$. We conclude with a discussion of noncommutative Poisson summation in more general compact Lie groups. As in the rest of the paper, the motivation for this result stems from path integrals \cite{Beauvillain2}: Poisson summation turns out to be crucial, in that context, to properly account for the topology of $G$ when deriving the mixed propagator $\langle p_f|e^{-iHt}|g_i\rangle$.

\subsection{The U(1) case; standard Poisson summation}
\label{ssec: U(1) case}

The U(1) group illustrates how compactness affects the Fourier construction through a projector \eqref{proji} on $\mathcal{I}$-invariant, \i{i.e.}~periodic, functions. The purpose of this example will therefore primarily be to show how to deal with formal infinite $|\mathcal I|$ factors.

\paragraph{Hilbert space $\boldsymbol L^{\boldsymbol2}$(U(1)).} Let us go through the ingredients of sections \ref{sec: operator algebra}-\ref{sec: representations} for the U(1) group. Its Lie algebra $\mathfrak{u}(1)\cong\mathbb{R}$ is trivially identified with the dual $\mathfrak{u}(1)^*$. The exponential map from $\mathfrak{u}(1)$ to U(1) is $\exp(X)=e^{iX}$, so the principal branch of the logarithm is the open interval $(-\pi,\pi)$, which is mapped by the exponential on $\text{U(1)}\backslash\{-1\}$. The Baker-Campbell-Hausdorff formula \eqref{baker} is trivial since the group is commutative, \i{i.e.}~$B(X,Y)=X+Y$. The elements \eqref{ix} that exponentiate to the identity are $\mathcal{I}=\{2\pi n\,|\,n\in \mathbb{Z}\}$.

Momenta commute, so no ordering prescription such as \eqref{eq: symmetric ordering} is required, and eq.~\eqref{expo} holds in any case for exponentials of momentum operators. The latter act on the Hilbert space $L^2(\text{U(1)})$ in the position representation \eqref{e20}, or on the space $L^2_{\star}(\mathfrak{u}(1)^*)\cong L^2(\mathbb{R})$ in the `large' momentum representation \eqref{eq: momentum op in momentum space}--\eqref{eq: zeta momentum rep}, in which case star products \eqref{eq: def star product} become pointwise multiplication. Since no nonzero function in $L^2(\mathbb{R})$ is periodic, the problem pointed out in \eqref{e38} remains: $e^{2\pi in\hat{p}}\cdot\Phi(p)=e^{2\pi inp}\Phi(p)\neq\Phi(p)$. The only way to fix this is to reduce the momentum representation as in section \ref{ssec: momentum rep 2}, using the projector \eqref{proji} which now reads
\begin{equation}
\label{s255}
\hat P_{\cal I}
=
\frac{1}{|\mathbb{Z}|}\sum_{n\in\mathbb{Z}}e^{2\pi in\hat{p}}.
\end{equation}
The Hilbert space $\hat P_{\cal I}\cdot L^2(\mathbb{R})$ thus consists of square-integrable $2\pi$-periodic functions, \i{i.e.}\ func\-tions on the circle, which is trivially isometric to $L^2(\text{U(1)})$. In that context, one may use $\int_{-\pi}^{\pi}\dd X = \frac{1}{|\mathbb{Z}|}\int_{-\infty}^{+\infty} \dd X$ to integrate functions on U(1) seen as $2\pi$-periodic functions on $\mathbb{R}$.

\paragraph{Fourier series.} Let us move now to the material of section \ref{sec:DefFourier} applied to U(1). First, the plane waves \eqref{e63} are just exponentials $E(X,p)=e^{-ipX}$. The Fourier transform \eqref{eq: def ncft}, which is commutative here, is the map $L^2(\mathbb{R})\to L^2(\mathbb{R})$ defined as usual by
\begin{align}
      \cF[\psi](p) 
      :=
      \int_{-\infty}^{\infty}\dd X \;e^{-ipX}\psi(X).
\end{align}
The Jacobian \eqref{e19} is trivial, and eqs.~\eqref{s15}--\eqref{e65} hold without any star product. As for the $\cal I$-invariant plane waves \eqref{s16bis}, they read
\begin{align}
     E_\mathcal{I}(\theta,p)
     =
     \frac{1}{|\mathbb{Z}|}\sum_{n\in \mathbb{Z}} e^{-ip(\theta+2\pi n)}=e^{-ip\theta}\sum_{k\in \mathbb Z}\frac{\delta(p-k)}{|\mathbb{Z}|},
     \label{e87}
\end{align}
which is a trivial special case of eq.~\eqref{e85}. Evaluating the plane wave \eqref{e87} at $p=n\in \mathbb{Z}$ yields $E_\mathcal{I}(\theta,n)=e^{-in\theta}$ which, considering the far right-hand side, justifies the prescription $\delta(0)=|\mathbb Z|$ (recall the discussion around eq.~\eqref{squaredelta}). In the end, one has
\begin{align}
    E_\mathcal{I}(\theta,p)
    =
    \begin{cases}
        e^{-in\theta} &\text{if } p=n\in \mathbb Z,
        \\
        0 &\text{otherwise.}
    \end{cases}
    \label{e89}
\end{align}
This is to say that $\cal I$-invariant plane waves \eqref{s16} are best seen as Fourier modes.

To obtain Fourier coefficients following section \ref{sec:DefFourier}, one imposes $\mathcal I$-invariance to the argument of the Fourier transform. From \eqref{eq: best def Fourier series} and the expression \eqref{e89} of periodic plane waves, Fourier coefficients read
\begin{align}
\label{e555}
    F_\mathcal{I}(\psi)(p)
    =
    \begin{cases}
    \sqrt{\delta(0)}\int_{-\pi}^\pi \dd X\; e^{-inX}\psi(X) &\text{if } p=n\in \mathbb Z,\\
    0 &\text{otherwise.}
    \end{cases}
\end{align}
This was to be expected: Fourier coefficients are supported on integers in momentum space $\mathfrak{u}(1)^*\cong\mathbb R$. The Fourier coefficients \eqref{e555} of a function $\psi\in L^2(\text{U}(1))$ may thus be written as
\begin{align}
    \cF_{\cal I}[\psi](p)
    =
    \sum_{n\in \mathbb Z}
\sqrt{\delta(p-n)}
\int_{-\pi}^\pi \dd X\; e^{-in X}\psi(X), 
\end{align}
where the square root of the delta distribution enforces enforces square-integrability of $ \cF_{\cal I}[\psi]$, similarly to eq.~\eqref{squaredelta}. Indeed, eq.~\eqref{e76} holds and may be recast as
\begin{align}
       \big\langle\cF_{\cal I}[\phi]\big|\cF_{\cal I}[\psi]\big\rangle_\mathbb{R}
       =
       \int \frac{\dd p}{2\pi}\sum_{n,m\in \mathbb{Z}}\bar{\phi}_n\psi_m\sqrt{\delta(p-m)\delta(p-n)}
       =
       \frac{1}{2\pi}\sum_{n\in \mathbb{Z}}\bar\phi_n\psi_n
\end{align}
in terms of standard Fourier coefficients $\phi_n:=\int_{-\pi}^{\pi}\dd X\,e^{inX}\phi(X)$ and similarly for $\psi$. This is to say that $\hat P_\mathcal{I}\cdot L^2(\mathbb R)\cong\ell^2(\mathbb{Z})\cong L^2(\text{U(1)})$, as had to be the case. Finally, the adjoint \eqref{defad} of Fourier coefficients is given by
\begin{align}
      \cF_{\mathcal{I}}^{\dagger}[\Phi](X) 
      =
      \sqrt{\delta(0)}\int \frac{\dd p}{{2\pi}}e^{ipX}\sum_{n\in \mathbb{Z}}\varphi_n\sqrt{\delta(p-n)}=\frac{1}{{2\pi}}\sum_{n\in\mathbb{Z}}\varphi_n e^{inX}.
\end{align}
This is the usual Fourier-series representation of a $2\pi$-periodic function $\psi(X)$.

\paragraph{Poisson summation.} The Poisson summation formula for U(1) is obtained by relating Fourier transforms and series through the projector \eqref{s255}, as follows. On the one hand, let $\psi(X)$ be a function in the `large' Hilbert space $L^2(\mathfrak{u}(1))=L^2(\mathbb R)$ of section \ref{ssec: position rep 2}. Acting on $\psi$ with the projector \eqref{s255} turns $\psi$ into a periodic function, namely $\sqrt{|\mathbb Z|}\hat P_\mathcal{I}\psi(x) = \frac{1}{\sqrt{|\mathbb Z|}}\sum_{n\in \mathbb Z} \psi(x+2\pi n)$. On the other hand, the commutative diagram \eqref{eq: commutative diagram of F} ensures that the Fourier transform and its inverse commute with the projector $\hat P_{\cal I}$. It follows that $\sqrt{|\mathbb Z|}\hat P_\mathcal{I}\psi(x) = \cF^\dagger_\mathcal{I}[\cF[\psi]](x)$, where $\cF[\psi]$ is the Fourier \i{transform} of $\psi$ while $\cF_{\cal I}^{\dagger}$ is the Fourier \i{series} \eqref{defad}. Since both computations of $\sqrt{|\mathbb Z|}\hat P_\mathcal{I}\psi(x)$ must give the same result, the definition \eqref{defad} of $\cF_{\cal I}^{\dagger}$ allows us to write
\begin{align}
    \sum_{n\in \mathbb Z} \psi(x+2\pi n) 
    =
    \int \frac{\dd p}{2\pi}\sum_{k\in \mathbb Z} \delta(p-k)e^{ipx}\cF[\psi](p) 
    =
    \frac{1}{2\pi}\sum_{k\in \mathbb Z} e^{ikx}\cF[\psi](k).
    \label{e95}
\end{align}
where we used the expression \eqref{e87} of $\mathcal{I}$-invariant plane waves.

Eq.~\eqref{e95} is the standard Poisson summation formula. The virtue of the derivation shown here is that it extends to any (compact) Lie group, as we now show.

\subsection{The SU(2) case; nonabelian Poisson summation}
\label{ssec: SU(2)}

The SU(2) group illustrates how both compactness and noncommutativity affect the construction of Fourier coefficients and series, so we discuss it in detail here. The presentation is organized as follows. We begin with a geometric preliminary on the principal branch and the Hilbert space $L^2(\text{SU}(2))$. Then we introduce Fourier modes and Fourier coefficients. In particular, we show that the latter simplify considerably in the case of class functions, and eventually recover the Kirillov character formula. We conclude with a new Poisson summation formula for functions in $\mathbb{R}^3\cong\mathfrak{su}(2)$.

\paragraph{Principal branch of SU(2).} The Lie algebra $\mathbb{R}^3\cong\mathfrak{su}(2)$ of SU(2) is the vector space $\mathbb{R}^3$ of traceless, Hermitian matrices.\footnote{Technically, the Lie algebra consists of \i{anti}-Hermitian matrices. These are just obtained by multiplying any Hermitian matrix by $i$.} A convenient basis (orthonormal with respect to the suitably normalized Killing form) is provided by Pauli matrices
\begin{equation}
      \sigma_x := 
      \begin{pmatrix}
          0 & 1
          \\
          1 & 0
      \end{pmatrix}
      ,\qquad
      \sigma_y := 
      \begin{pmatrix}
          0 & -i
          \\
          i & 0
      \end{pmatrix}
      ,\qquad
      \sigma_z :=
      \begin{pmatrix}
          1 & 0
          \\
          0 & -1
      \end{pmatrix}.
\end{equation}
Any element of $\mathfrak{su}(2)$ can thus be written as a linear combination $X = X^j\sigma_j = \vec X\cdot \vec \sigma$, where $\vec X=(X^x,X^y,X^z)\in\mathbb{R}^3$. The exponential map from $\mathfrak{su}(2)$ to SU(2) reads
\begin{align}
      \exp(iX^j\sigma_j)
      =
      e^{i\vec X \cdot \vec \sigma} 
      =
      \mathbb{I}\cos\|\vec X\| + i \,\vec u_X\cdot \vec\sigma\sin\|\vec X\|,
      \label{eq: exp su2}
\end{align}
where $\vec u_X := \vec X/\|\vec X \|$ denotes the unit vector pointing in the direction of $\vec X$ and $\|\vec X\|:=\sqrt{X^iX^i}$ is the usual Euclidean norm.

The principal branch of the logarithm, on which the exponential \eqref{eq: exp su2} is injective with a dense image in SU(2), is the open ball of radius $\pi$ centred at the origin in $\mathfrak{su}(2)\cong\mathbb{R}^3$. See fig.~\ref{fig:principalbranch} for a cartoon, and recall the right panel of fig.~\ref{fig:logs} for the actual picture. The only missing element of SU(2) in the image of the principal branch is $-\mathbb{I}$, which can be obtained as the exponential of any point on the sphere of radius $\pi$ in $\mathfrak{su}(2)$. As readily seen from \eqref{eq: exp su2}, the set $\mathcal{I}$ of elements that exponentiate to the identity are vectors of integer radius:
\begin{align}
{\cal I}
=
\operatorname{Logs}(e)
=
\big\{\vec X \in \mathbb{R}^3\,\big|\,\|\vec X\| \in 2\pi\mathbb{N}\big\}.
\end{align}
This is the expected union \eqref{eq: sum over I unpacked} of adjoint orbits, which are spheres in this case; see again the right panel of fig.~\ref{fig:logs}. We stress that the logarithms of a generic Lie algebra element are far less numerous: provided $\|\vec X\|\notin2\pi\mathbb{N}$,
\begin{align}
\label{e105}
    \operatorname{Logs}\big(\exp(X)\big)
    =
    \{\vec X + 2\pi n \vec u_X\,|\, n\in \mathbb{Z}\}.
\end{align}
This is the rank-one version of eq.~\eqref{e84}. Therefore, each branch of the logarithm is labelled by an element of $\mathbb Z$.

\paragraph{Hilbert space $\boldsymbol L^{\boldsymbol2}$(SU(2)).} As in section \ref{ssec: position rep 1}, consider the space of wave functions on SU(2) that are square-integrable with respect to the Haar measure $\dd g$. In terms of the vector $\vec X$ in eq.~\eqref{eq: exp su2}, the Jacobian \eqref{e19} between the flat measure on $\mathbb{R}^3$ and the Haar measure reads \cite{Taylor} (see also \cite{Helgason})
\begin{align}
\label{jack}
      J(\vec X) 
      =
      \frac{\sin^2\|\vec X\|}{\|\vec X\|^2}.
\end{align}
The scalar product \eqref{e18} can thus be recast as
\begin{align}
      \int_G \dd g\, \bar\phi(g)\psi(g)
      =
      \int\limits_{\|\vec X\|< \pi}\frac{\sin^2\|\vec X\|}{\|\vec X\|^2} \dd^3 X\,\bar\phi(X)\psi(X)
\end{align}
in principal branch coordinates. Similarly, the Fourier transform \eqref{eq: def ncft} of a function $\psi\in L^2(\mathfrak{su}(2))=L^2(\mathbb{R}^3)$ reads
\begin{align}
    \cF[\psi](p) 
    =
    \int_{\mathbb R^3} \frac{\sin^2\|\vec X\|}{\|\vec X\|^2}\dd ^3 X\, e^{-i\langle p,\vec X\rangle}\psi(X).
    \label{e100}
\end{align}
This will be crucial for Poisson summation, including the effects of the Jacobian \eqref{jack}.

\begin{figure}
\centering
\includegraphics[width=.8\textwidth]{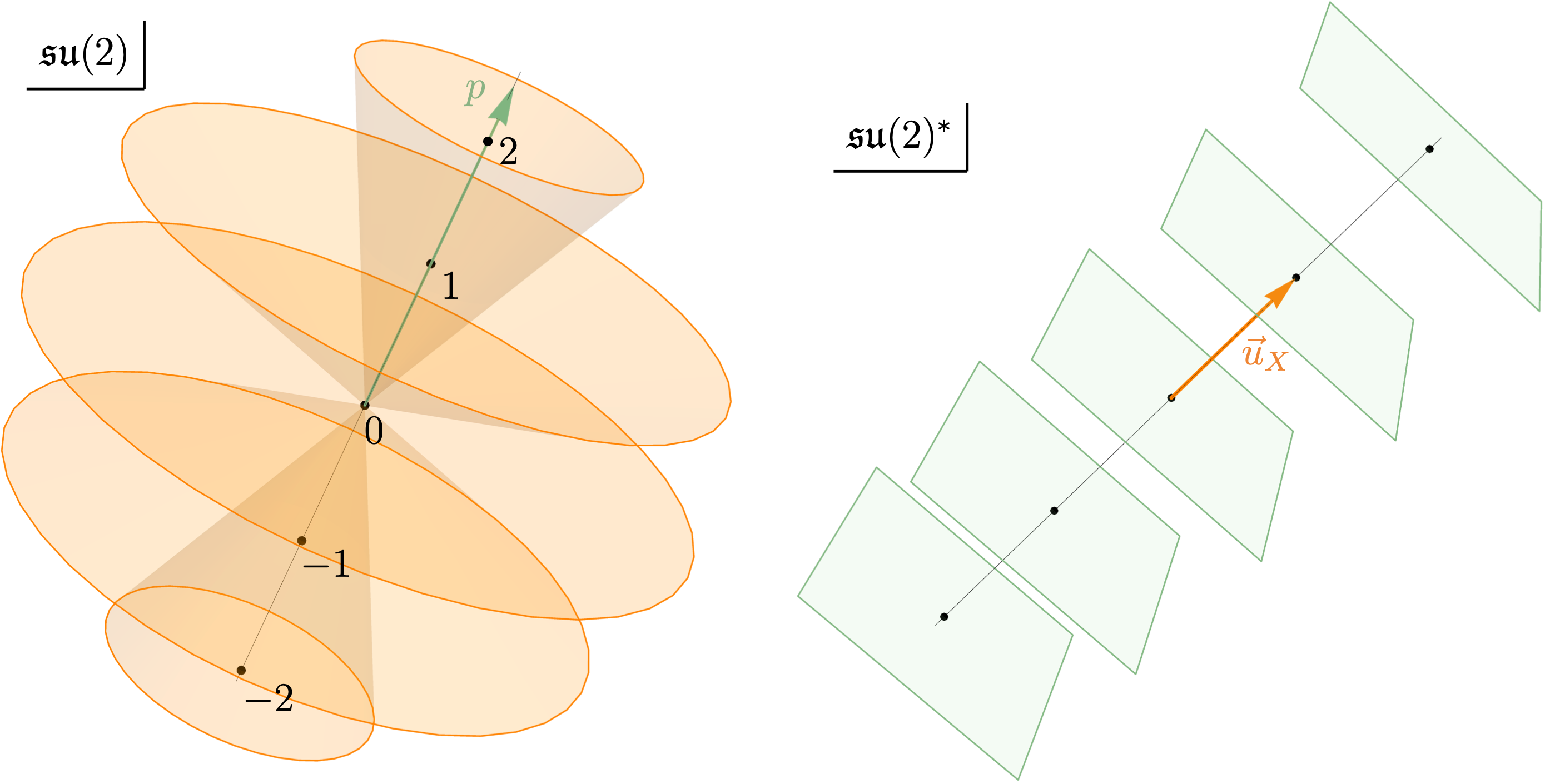}
    \caption{Localization of ${\cal I}$-invariant plane waves \eqref{e103} in $\vec X$ at fixed $p$ (left), and in $p$ at fixed $\vec X$ (right). On the left panel, we chose $\|p\|=2.2$ so that only five values of $m$ are allowed in \eqref{e103}. On the right panel, the plane waves are localized on planes perpendicular to $\vec u_X$ that are evenly spaced by integer multiples of $\vec u_X$.}
    \label{fig:localization}
\end{figure}

\paragraph{Fourier modes.} Let us now turn to the $\cal I$-invariant plane waves \eqref{s16bis}. Since the set of logarithms is given by \eqref{e105} for a generic Lie algebra element $X$, one can write 
\begin{align}
    E_\mathcal{I}(\vec X,p) 
    =
    \frac{1}{|\mathbb Z|}\sum_{n\in \mathbb{ Z}}e^{-i\langle p, \vec X + 2\pi n \vec u_X\rangle}= e^{-i\langle p,\vec X\rangle}\sum_{m\in \mathbb Z} \frac{\delta(\langle p,\vec u_X\rangle-m)}{|\mathbb Z|}
    \label{e103}
\end{align}
for almost any $\vec X$. This is the rank-one version of eq.~\eqref{e85}. Each such plane wave localizes on those surfaces where $\langle p,\vec u_X\rangle$ is an integer, both when seen as a function of $p$ at fixed $X$, and as a function of $X$ at fixed $p$. See fig.~\ref{fig:localization}.

Let us unpack the condition $\langle p,\vec u_X\rangle\in\mathbb{Z}$. At fixed $p$ and varying $\vec X$, the Cauchy-Schwarz inequality guarantees $|\langle p,\vec u_X\rangle| \leq \|p\|$. Hence there are at most $2\lfloor \|p\|\rfloor +1$ non-vanishing terms on the right-hand side of \eqref{e103}, namely those for which $m\in \{-\lfloor \|p\|\rfloor,\dots, \lfloor\|p\|\rfloor\}$. Let us therefore give the suggestive name $j(p) := \lfloor \|p\|\rfloor$, thought of as the (integer) spin of a representation of SU(2) obtained by geometric quantization of a coadjoint orbit whose radius is fixed by $p$. In that case, $\vec u_X$ is localized on circles of constant integer scalar product with $p$, and $\vec X$ is localized on half-cones passing through these circles; see the left panel of fig.~\ref{fig:localization}. Conversely, at fixed $\vec X$, $p$ can be split as  $p=p_\parallel + p_\perp$, such that $\langle p, \vec u_X\rangle = p_\parallel$. Then $E_\mathcal{I}(\vec X, p)$ is localized on affine planes perpendicular to $\vec u_X$, passing through $m\vec u_X$. We will call these planes $\mathcal P_m(\vec X)$; see the right panel of fig.~\ref{fig:localization}. Up to a multiplicative constant, these are planes perpendicular to the root lattice adapted to $\vec X$.

\paragraph{Fourier coefficients for SU(2).} The definition \eqref{eq: best def Fourier series} of Fourier coefficients, applied to the SU(2) case in \eqref{e100}, yields
\begin{align}
      \cF_{\cal I}[\psi](p)
      =
        \sqrt{\mathbb Z}\int_{\|\vec X\|<\pi}\dd^3 X\,\frac{\sin^2\|\vec X\|}{\|\vec X\|^2}\,\psi(X)\!\!\!\sum_{m=-j(p)}^{j(p)} e^{-im\|\vec X\|}\frac{\delta(\langle p,\vec u_X\rangle -m)}{|\mathbb{Z}|}
\end{align}
for any wave function $\psi\in L^2(\text{SU(2)})$, where we used the $\cal I$-invariant plane wave \eqref{e103}. To simplify this expression, let us work in spherical coordinates $(r,\theta,\varphi)$ adapted to $p$, such that the north pole  $\theta=0$ is aligned with $p$.\footnote{For SU(2), spherical coordinates naturally arise when using the Weyl integration formula that splits integrals in $\mathfrak{su}(2)$ into toroidal (radial) directions and adjoint-orbit (angular) directions.} Then the noncommutative Fourier coefficients of $\psi$ are
\begin{equation}
\begin{split}
    &\cF_{\cal I}[\psi](p)=\\
    &\frac{1}{\sqrt{|\mathbb{Z}|}}
    \sum_{m=-j(p)}^{j(p)}
    \int_0^\pi \dd r\,\sin^2(r) e^{-imr}
    \int_0^\pi\sin(\theta)\dd \theta\, \delta\big(\|p\|\cos(\theta) -m\big)
    \int_0^{2\pi}\dd \varphi\,\psi_p(r,\theta,\varphi),
    \end{split}
\end{equation}
where the notation $\psi_p(r,\theta,\varphi)$ stresses that the argument of $\psi$ depends on $p$. Changing the integration variable from $\theta$ to $u=\|p\|\cos(\theta)$ makes it possible to integrate the delta function, leading to
\begin{align}
    \cF_{\cal I}[\psi](p)
    =
    \frac{1}{\sqrt{|\mathbb{Z}|}}
    \frac{1}{\|p\|}\sum_{m=-j(p)}^{j(p)}\int_{0}^\pi \dd r\; \sin^2(r) e^{-imr}
    \int_0^{2\pi}\dd \varphi\; \psi_p\big(r,\arccos(m/\|p\|),\varphi\big).
    \label{e106}
\end{align}
No further simplification is available, in general, for arbitrary functions $\psi$.

\paragraph{Class functions.} By definition, a class function is a function $\psi(g)$ which is constant on conjugacy classes, meaning that $\psi(h g h^{-1}) = \psi(g)$ for all $g,h\in G$. The integral \eqref{e106} simplifies drastically for any such function. In the SU(2) case, class functions only depend on the radial coordinate $r$, so the sum over $m$ and the integral over $\varphi$ can be carried out explicitly. The result is
\begin{align}
    \cF_{\cal I}[\psi](p)
    =
    \frac{1}{\sqrt{|\mathbb{Z}|}}
    \frac{4\pi}{\|p\|}\int_0^\pi\dd r\; \sin(r)\cos(r/2)\sin\big((j(p)+1/2)r\big)\psi(r)
    \text{~~~($\psi$ a class function).}
    \label{e107}
\end{align}
A basis of class functions is provided by the characters of irreducible representations. Let us therefore compute the Fourier transform of the character of a highest-weight representation of SU(2)---say $\rho_{\lambda}$, with spin $\lambda$, in the Hilbert space $\mathbb C^{2\lambda+1}$. In principal branch coordinates, the character of the representation reads
\begin{align}
    \chi_\lambda (\vec X)
    := 
    \operatorname{tr}\Big[\rho_\lambda\big(\exp(X)\big)\Big]
    =
    \frac{\sin\big((2\lambda+1)\|\vec X\|\big)}{\sin\|\vec X\|}.
    \label{e110}
\end{align}
Plugging this expression into \eqref{e107} yields the Fourier transform
\begin{align}
    \cF_\mathcal{I}[\chi_\lambda](p)
    =
    \begin{cases}
        \frac{\pi^2}{\sqrt{|\mathbb{Z}|}}\frac{1}{\|p\|} & \text{if } 2\lambda \leq \|p\|<2\lambda+2,
        \\
        0 & \text{otherwise.}
    \end{cases}
    \label{e109}
\end{align}
In particular, this exhibits the fact that the Fourier transforms of different characters are mutually orthogonal.

For consistency, one may also check that the Fourier series given by eq.~\eqref{e109} gives back the character \eqref{e110}. This can be done using the lone star lemma \eqref{eq: no star inv Fourier} with $r=1$, as follows:
\begin{align}
    F_{\cal I}^\dagger\big[\cF_\mathcal{I}[\chi_\lambda]\big](\vec X) 
    &=
    \sqrt{|\mathbb Z|}\frac{\|\vec X\|^2}{\sin^2(\|\vec X\|)}\int \frac{\dd^3 p}{(2\pi)^3}\,\cF_\mathcal{I}[\chi_\lambda](p) e^{i\langle p, \vec X\rangle}
    \label{e524}
    \\
    &= 
    \frac{\|X\|^2}{4\sin^2(\|X\|)} \int_{2\lambda}^{2\lambda +2} \!\!p\,\dd p \int_0^\pi \sin \theta\,\dd \theta \,e^{ip\|X\|\cos\theta}
    \notag
    \\
    &=
    \frac{\text{sin}((2\lambda +1)\|\vec X\|)}{\sin\|\vec X\|}
    =
    \chi_\lambda(\vec X),
    \label{e525}
\end{align}
where we chose to work in spherical coordinates adapted to $p$.

It is tempting to relate eq.~\eqref{e107} to the Parseval-Plancherel identity. The latter states, in the SU(2) case, that the Fourier transform
\begin{equation}
    \hat \psi(\lambda) 
    :=
    \int \dd g\; \psi(g) \bar \chi_\lambda(g) = 4\pi \int \dd r\; \sin(r)\sin((2\lambda + 1) r) \psi(r)
    \label{e107bis}
\end{equation}
yields an isometry between the space of class functions (with the $L^2(G)$ norm) and the space of functions on the Pontryagin dual of $G$, $\hat G$, with norm
\begin{equation}
\label{planche}
    \|\hat \psi\|^2 
    =
    \sum_{\lambda \in \hat G} \dim(\lambda) \overline{\hat\psi(\lambda)}\hat \psi(\lambda).
\end{equation}
In the case at hand, the Fourier transform \eqref{e107} does look similar to eq.~\eqref{e107bis}, but it differs from it at the end of the day. The difference was to be expected, since momentum space is discrete in one case, while it is continuous in the other. Remarkably, we will see in section \ref{Duflo map} that the noncommutative Fourier coefficients associated with Duflo quantization actually reproduce the Parseval-Plancherel identity.

\paragraph{Kirillov character formula.} Note that the Fourier coefficients \eqref{e109} of a character are localized around the coadjoint orbit entering Kirillov's character formula \cite{KirillovLectures}, namely the sphere with radius $\|p\|=2\lambda +1$. The derivation \eqref{e524}--\eqref{e525} can in fact be understood as a blurred-out version of the character formula, where one integrates over a family of spheres rather than a single sphere of radius $2\lambda +1$. In fact, eq.~\eqref{e109} can be directly related to the Kirillov character formula thanks to a pseudodifferential operator trick, as follows.

Let us first focus on the distributional part of \eqref{e109}. For convenience, let $h(r):=\Theta(r- 2\lambda) - \Theta(r-(2\lambda+2))$ where $\Theta$ denotes the Heaviside distribution, so that the Fourier coefficients \eqref{e109} read $F_\mathcal{I}[\chi_\lambda](p) = \frac{\pi^2}{\sqrt{|\mathbb Z|}}\frac{1}{\|p\|}h(\|p\|)$. The derivative of $h$ is $h'(r)=\delta(r-2\lambda) - \delta(r-(2\lambda+2))$. The idea now is to relate this function to one that is fully localized at $2\lambda +1$. To that end, note that $\delta(r-(2\lambda +1 \pm 1)) = e^{\pm\partial_r}\delta(r-(2\lambda +1))$, so that
\begin{align}
    h'(r) = \left(e^{\partial_r}-e^{-\partial_r}\right)\delta(r-(2\lambda +1)) = 2\sinh(\partial_r)\delta(r-(2\lambda +1)).
\end{align}
In pseudodifferential operator language, integrating amounts to multiplying by $1/\partial_r$, so that
\begin{align}
    h(r) 
    =
    \frac{2\sinh(\partial_r)}{\partial_r}\delta(r-(2\lambda+1)) = 2\operatorname{sinc}(i\partial_r)\delta(r-(2\lambda+1)).
\end{align}
Because $\operatorname{sinc}(r)$ is an even function, its Taylor expansion only contains even powers $r$, \i{i.e.}~integer powers of $r^2$. This means that the differential operator $\operatorname{sinc}(i\partial_r)$ can be expressed as a series of iterated Laplacians. On radial functions (class functions), powers of the Laplacian $\Delta$ act as
\begin{equation}
    \Delta^k\left(\frac{\Psi(r)}{r}\right) = \frac{\Psi^{(2k)}(r)}{r},
\end{equation}
so the Fourier coefficients \eqref{e109} can be recast as
\begin{align}
    F_\mathcal{I}[\chi_\lambda](p) 
    =
    \frac{2\pi^2}{\sqrt{|\mathbb Z|}} \operatorname{sinc}\!\left(i\frac{\partial}{\partial p}\right) \frac{\delta(\|p\|-(2\lambda+1))}{\|p\|}.
    \label{eBANGER ABSOLU}
\end{align}
The Fourier series that corresponds to this is nothing but the Kirillov character formula. Indeed, eq.~\eqref{eq: no star inv Fourier} yields
\begin{align}
    \chi_\lambda(\vec X)
=    
    F^\dagger_\mathcal I[F_\mathcal{I}[\chi_\lambda]](\vec X) 
    \stackrel{\text{\eqref{eBANGER ABSOLU}}}{=}
    2\pi^2\frac{\|\vec X\|^2}{\sin^2 \|\vec X\|}\int\frac{\dd^3 p}{(2\pi)^3}e^{i\langle p, \vec X\rangle} \operatorname{sinc}\!\left(i\frac{\partial}{\partial p}\right)\frac{\delta(\|p\|-(2\lambda +1))}{\|p\|}.
\end{align}
Integrating by parts here and using the fact that $-i\partial/\partial p$ is multiplication by $\vec X$ when it acts on plane waves, the pseudodifferential operator simplifies with half of the Jacobian prefactor. The integral on the right-hand side then fully localizes on the sphere of radius $2\lambda +1$, which again is the precise coadjoint orbit appearing in the Kirillov character formula. The final expression reads
\begin{align}
    \chi_\lambda(\vec X) 
    =
    \frac{1}{4\pi}\frac{\|\vec X\|}{\sin\|\vec X\|}\int (2\lambda + 1) \sin \theta \dd\theta \dd \varphi\; e^{i(2\lambda +1)\|\vec X\|\cos\theta},
\end{align}
where one recognizes the Liouville volume form $\mu\sin\theta\dd \theta \dd \varphi$ on the SU(2) coadjoint orbit (the sphere) of radius $\mu$.

We conjecture that such a relation holds quite generally, for any compact Lie group, between noncommutative Fourier coefficients of characters and the Kirillov character formula. More precisely, one may expect the noncommutative Fourier transform of an irreducible character of a compact group $G$ to take the form \eqref{e109}, with a blurred-out localization around the relevant coadjoint orbit. (Specifically, eq.~\eqref{e109} is consistent with an integration over the family of orbits $\{\mathcal{O}_{\lambda + c^ih_i}$ with parameters $c_i\in [0,1)$ and $i \in \{1,\dots r\}\}$, where $h_i$ denotes the positive roots of $\mathfrak{g}$.) We will not attempt to derive such a general result here.

\paragraph{Poisson summation for SU(2).} Let $\psi$ be a complex function on $\mathfrak{su}(2)\cong\mathbb{R}^3$ that falls off sufficiently fast at infinity. In certain computations involving path integrals for a quantum rigid body \cite{Beauvillain2}, the following result turns out to be essential:

\noindent\i{Statement.} For almost any $X\in\mathfrak{su}(2)$, one has the Poisson summation formula
\begin{align}
    \sum_{n\in\mathbb Z}\psi(\vec X + 2\pi n\vec u_X) 
    =
    \frac{-1}{\sin^2\|\vec X\|}\sum_{m\in \mathbb Z} e^{im\|\vec X\|}\int_{\mathcal P_m(\vec X)}\frac{\dd^2 p}{(2\pi)^3}\partial_{\vec u_X}^2\cF[\psi](p).
    \label{e114}
\end{align}
Here $\mathcal P_m(\vec X)$ denotes the affine plane in $\mathbb{R}^3$ perpendicular to $\vec X$ and going through $m \vec u_X$ (see Fig.~\ref{fig:localization}), $\partial_{\vec u_X}$ denotes the partial derivative in the $\vec u_X$ direction, and $\cF[\psi]$ is the Fourier transform \eqref{e100}.
\begin{center}
\begin{minipage}{.9\textwidth}
{\footnotesize%
\i{Proof of eq.~\eqref{e114}.}
As explained under \eqref{e85}, deriving the Poisson summation formula for a function $\psi\in L^2(\mathfrak{su}(2))$ amounts to equating $\sqrt{|\mathbb Z|}\hat P_\mathcal{I}\psi$ and $F_\mathcal{I}^\dagger(F(\psi))(X)$. On the one hand, eq.~\eqref{big nctranslation} used in the projector \eqref{proji} implies
\begin{align}
\sqrt{|\mathbb Z|}
    \hat P_\mathcal{I}\psi(\vec X)
    =
    \frac{1}{\sqrt{|\mathbb Z|}}\sum_{n\in\mathbb Z}\psi(\vec X + 2\pi n\vec u_X)
    \label{s30}
\end{align}
for almost any $X\in\mathfrak{su}(2)$. On the other hand, let $\cF[\psi] := \Psi$ be the Fourier transform \eqref{e100} of $\psi$. The inverse Fourier transform in its form \eqref{eq: no star inv Fourier} yields
\begin{align*}
    \hat P_\mathcal{I}F^\dagger(\Psi)(\vec X) 
    =
    \frac{1}{\sqrt{|\mathbb Z|}}\sum_{m\in \mathbb Z}\int \frac{\dd^3 p}{(2\pi)^3}\, \frac{(\vec X + 2\pi m \vec u_X)^2}{\sin^2\|\vec X\|}e^{i\langle p, \vec X +2\pi m\vec u_X\rangle}\Psi(p),
\end{align*}
where the factor $\sim\sin^2\|\vec X\|/\|\vec X\|^2$ is the Jacobian \eqref{jack} of the Haar measure. Now integrating by parts and rewriting $(\vec X + 2\pi m \vec u_X)^2e^{i\langle p, \vec X +2\pi m\vec u_X\rangle}$ as a Laplacian $\Delta_p(-e^{i\langle p, \vec X +2\pi m\vec u_X\rangle})$, one obtains
\begin{align*}
    \sqrt{|\mathbb Z|}\hat P_\mathcal{I}F^\dagger(\Psi)(\vec X)
    =
    -
    \frac{1}{\sqrt{|\mathbb Z|}\sin^2\|\vec X\|}\int \frac{\dd^3 p}{(2\pi)^3}\, \Delta_p\Psi(p)\sum_{m\in \mathbb Z} e^{i\langle p, \vec X +2\pi m\vec u_X\rangle}.
\end{align*}
Finally using eq.~\eqref{e103} and the localization of plane waves discussed there, one can further simplify the result: if $\mathcal P_m(\vec X)$ denotes the affine plane perpendicular to $\vec X$ through $m \vec u_X$, and if $\partial_{\vec u_X}$ is the partial derivative in the $\vec u_X$ direction, then
\begin{align*}
    \hat P_\mathcal{I}F^\dagger(\Psi)(\vec X)
    =
    -
    \frac{1}{\sqrt{|\mathbb Z|}\sin^2\|\vec X\|}\sum_{m\in \mathbb Z} e^{im\|\vec X\|}\int_{\mathcal P_m(\vec X)}\frac{\dd^2 p}{(2\pi)^3}\partial_{\vec u_X}^2\Psi(p).
\end{align*}
Equating this with \eqref{s30} leads to the Poisson summation formula \eqref{e114}.
\hfill$\blacksquare$}
\end{minipage}
\end{center}

\subsection{Poisson summation for any group}
\label{ssec: poisson}

We now sketch how the derivation of a Poisson summation formula extends to generic compact Lie groups. Eqs.~\eqref{e95} and \eqref{e114} are thus special cases of a more general result on noncommutative Fourier series. As mentioned in section \ref{ssequant}, this is typically crucial for path integral computations in the Hilbert space $L^2(G)$ \cite{Beauvillain2}.

Note first that the SU(2) Poisson summation formula \eqref{e114} involves two-dimensional integrals. Indeed, for SU(2), the localization of $\mathcal I$-invariant plane waves \eqref{e103} occurs on planes rather than points, leading to integrals along those planes. For a more general Lie group $G$, Poisson summation similarly leads to  $(n-r)$-dimensional integrals, where $n=\dim(G)$ and $r$ is the rank (the dimension of a maximal torus). The idea of the derivation is thus the same as in SU(2): start with a function $\psi\in L^2(\mathfrak{g})$, then make it $\mathcal I$-invariant by acting with the projector \eqref{proji}. With the notation introduced around \eqref{e84}, and using the action \eqref{big nctranslation} of exponential operators on functions, this yields the periodic function
\begin{align}
    \hat P_\mathcal{I}\psi(X) 
    =
    \frac{1}{|\mathbb Z|^r}\sum_{\mathbf n\in \mathbb Z^r}\psi\big(X+2\pi n^ia_i(X)\big)
\end{align}
for almost any $X\in\mathfrak{g}$.

On the other hand, one can compute the Fourier transform $\Psi(p)=F(\psi)(p)$ given by \eqref{eq: def ncft}, and note that $\hat P_\mathcal{I}\psi = \hat P_\mathcal{I}F^\dagger(\Psi)=F^\dagger_\mathcal I(\Psi)/\sqrt{|\mathbb Z|^r}$. From the definition \eqref{defad}, the lone star lemma \eqref{eq: lone star lemma} and the expression \eqref{e85} of noncommutative plane waves, this yields (for almost any $X$)
\begin{align}
    \frac{1}{\sqrt{|\mathbb Z|^r}}F^\dagger_\mathcal I(\Psi) 
    =
    \frac{1}{|\mathbb Z|^r}\sum_{\mathbf n\in \mathbb Z^r}\frac{1}{J(X+2\pi n^ia_i(X))}\int_{\mathfrak{g}^*}\frac{\dd^np}{(2\pi)^n}e^{i\langle p, X+2\pi n^i a_i(X)\rangle}\Psi(p).
    \label{e116}
\end{align}
Now a key point, motivated by eq.~\eqref{e19}, is that $J(X)$ can be expressed as a fraction $J(X)=N(X)/D(X)$, where the numerator $N(X)$ is periodic under translations by $2\pi a_i(X)$ and the denominator $D(X)$ is polynomial in $X$. These are respectively $\sin^2\|X\|$ and $\|X\|^2$ for SU(2), owing to eq.~\eqref{jack}. Assuming this holds, the numerator can be put in front of the sum in \eqref{e116}, while the denominator can be recast as a differential operator $D(-i\partial_p)$ acting on plane waves, leading to
\begin{align}
    \frac{1}{\sqrt{|\mathcal I|}}F^\dagger_\mathcal I(\Psi) 
    =
    \frac{1}{|\mathbb Z|^rN(X)}\sum_{\mathbf n\in \mathbb Z^r}\int_{\mathfrak{g}^*}\frac{\dd^np}{(2\pi)^n}\left(D(-i\partial_p)e^{i\langle p, X+2\pi n^i a_i(X)\rangle}\right)\Psi(p).
\end{align}
Integration by parts here makes all derivatives hit on $\Psi$, and the sum of plane waves can be localized thanks to \eqref{e85}, so that
\begin{align}
\label{e118}
    \frac{1}{\sqrt{|\mathbb Z|^r}}F^\dagger_\mathcal I(\Psi) 
    =
    \frac{1}{|\mathbb Z|^rN(X)}\sum_{k\in \mathbb Z^r} \int_{\mathfrak{g}^*}\frac{\dd^np}{(2\pi)^n}e^{i\langle p, X\rangle}\left(\prod_{j=1}^r\delta(\langle p, a_j(X)\rangle-k_i)\right)D(i\partial_p)\Psi(p).
\end{align}
Now recall that the basis $\{a_i(X)\}$ of the Lie algebra $\mathfrak{t}_X$ of the maximal torus $T_X$ passing through $X$ is chosen so that $X\propto a_1(X)$; denote the proportionality constant by $\kappa(X)$. Then the exponential can be pulled out of the integral \eqref{e118}, yielding
\begin{align}
\label{e119}
    \frac{1}{\sqrt{|\mathbb Z|^r}}F^\dagger_\mathcal I(\Psi) 
    =
    \frac{1}{|\mathbb Z|^rN(X)}\sum_{k\in \mathbb Z^r} e^{ik_1\kappa(X)}\int_{\mathfrak{g}^*}\frac{\dd^np}{(2\pi)^n}\left(\prod_{j=1}^r\delta(\langle p, a_j(X)\rangle-k_i)\right)D(i\partial_p)\Psi(p).
\end{align}
The final step is to localize the integral of $\Psi$. To this end, let us explicitly split $\mathfrak{g}$ as a direct sum of a Cartan subalgebra $\mathfrak{t}_X$ and root spaces $\mathfrak{g}_{\alpha,X}$, labelled by $\alpha$: $\mathfrak{g}=\mathfrak{t}_X \oplus_\alpha \mathfrak{g}_{\alpha,X}$. One can accordingly split $\mathfrak{g}^*=\mathfrak{t}_X^* \oplus_\alpha \mathfrak{g}_{\alpha,X}^*$, so that $p = p_\parallel + p_\perp$, with $p_\parallel\in \mathfrak{t}_X^*$ and $p_\perp \in \oplus_\alpha\mathfrak{g}_{\alpha,X}^*$. The localization condition $\langle p, a_i(X)\rangle\in \mathbb Z$ then translates into the fact that $p_\parallel$ must belong to the root lattice of the Cartan subalgebra $\mathfrak{t}_X$. By contrast, $p_\perp$ is free, so only an integral in the direction of $p_\perp$ remains. Thus, in the differential operator $D(i\partial_p)$, all derivatives in the $\oplus_\alpha\mathfrak{g}_{\alpha,X}$ directions are total derivatives in the integral \eqref{e119}, to which they do not contribute for functions $\psi$ that decay at infinity. Let us therefore denote by $O(i\partial_{p_\parallel})$ the differential operator obtained from $D(i\partial_p)$ by discarding derivatives in the $\oplus_\alpha\mathfrak{g}_{\alpha,X}$ directions. All in all, eq.~\eqref{e119} becomes
\begin{align}
    \frac{1}{\sqrt{|\mathbb Z|^r}}F^\dagger_\mathcal I(\Psi)
    =
    \frac{1}{|\mathbb Z|^rN(X)}\sum_{\mathbf k\in \mathbb Z^r}e^{i\kappa(X)k_1} \int_{\oplus_\alpha \mathfrak{g}_{\alpha,X}^*}\frac{\dd^{n-r}p_\perp}{(2\pi)^n}O(i\partial_{p_{\parallel}})\Psi(\mathbf k, p_\perp),
\end{align}
from which we conclude the following:

\noindent\i{Statement.} Given a complex function $\psi$ on $\mathfrak{g}\cong\mathbb{R}^n$ that falls off sufficiently fast at infinity, one has the Poisson summation formula
\begin{align}
    \sum_{\mathbf n\in \mathbb Z^r}\psi(X+2\pi n^ia_i(X)) 
    =
    \frac{1}{N(X)}\sum_{\mathbf k\in \mathbb Z^r}e^{i\kappa(X)k_1} \int_{\oplus_\alpha \mathfrak{g}_{\alpha,X}^*}\frac{\dd^{n-r}p_\perp}{(2\pi)^n}O(i\partial_{p_{\parallel}})\Psi(\mathbf k, p_\perp)
    \label{poissum}
\end{align}
for almost any $X\in\mathfrak{g}$. Here the numerator $N(X)$ and the operator $O(i\partial_{p_{\parallel}})$ stem from the Jacobian \eqref{e19} as explained below eqs.~\eqref{e116} and \eqref{e119}.

To conclude, we stress once again the importance of all the results above when attempting to compute, in practice, any path integral of the form \eqref{pint}. Poisson summation, in particular, is unavoidable when the group manifold $G$ has compact dimensions. If these compact dimensions fail to commute with some of the generators of $G$, standard Abelian Poisson summation does not suffice, and one needs instead to use the full power of formulas such as eq.~\eqref{poissum}. All this and more will be explored in the separate paper \cite{Beauvillain2}.

\section{Noncommutative Fourier series with Duflo ordering}
\label{Duflo map}

In sections \ref{sec: operator algebra}--\ref{sec: examples}, we chose to use the symmetric quantization map \eqref{eq: symmetric ordering} to order the $p_i$s. This was motivated by the simple correspondence it induces between exponential functions and exponential operators \eqref{expo}. However, another common choice of quantization map for functions in $C^\infty(\mathfrak{g}^*)$ is the so-called Duflo map \cite{Duflo}. The latter does not map exponential functions to exponential operators, and leads to an alternative definition of noncommutative Fourier transforms and series.

In this short section, we describe the adaptation of all the material above to such a different choice of ordering of momentum operators. Most of the results carry over up to straightforward modifications of the formulas, involving \i{e.g.}~extra Jacobian factors (or, to the contrary, fewer Jacobian factors, hence simpler expressions). The most notable results occur at the very end: we show for instance that the Parseval-Plancherel formula straightforwardly follows from the Duflo-ordered noncommutative Fourier transform, and also revisit the Kirillov character formula. The latter shows that the Duflo-ordered Fourier coefficient of a character is localized on a single coadjoint orbit, which should be contrasted with the coefficient \eqref{e109} found with symmetric ordering.

\paragraph{Duflo map.} The operator algebra $\mathcal A_\text{mom}$, obtained by quantizing functions of $p$, is (isomorphic to the closure of) the universal enveloping algebra ${\cal U}(\mathfrak{g})$. A subalgebra of ${\cal U}(\mathfrak{g})$ is of particular interest: the \i{Casimir operators}, which commute with $\mathfrak{g}$. Relatedly, in the algebra of functions of $p$, a distinguished Poisson subalgebra consists of functions that are invariant under the coadjoint action, \i{i.e.}~\i{Casimir functions} such that $A(p) = A(\mathrm{Ad}_g^* p)$ for all $g\in G$. The key interest of Duflo's quantization map is that it provides an algebra isomorphism between Casimir operators and Casimir functions \cite{Duflo}.

In order to define the Duflo map, first introduce the function
\begin{align}
    j(X) 
    :=
    \det\left(\frac{\sinh(\mathrm{ad}_{X/2})}{\mathrm{ad}_{X/2}}\right),
\end{align}
related to the Jacobian of the exponential by $J(X) = j(X) e^{-\operatorname{tr} \mathrm{ad}_{X/2}}$. Note that the two coincide in the case of unimodular groups, since in that case $\tr \mathrm{ad} = 0$. The Duflo map is then defined as
\begin{align}
\label{duffing}
    \mathcal{D}(A(p))
    =
    \mathcal{Q}_\text{mom}\left(j^{1/2}(i\partial_p)A(p)\right)
\end{align}
for any function $A$ in momentum space. Here $\mathcal{Q}_\text{mom}$ denotes the symmetric ordering \eqref{eq: symmetric ordering} and, as in eq.~\eqref{eq: lone star lemma}, $f(\partial)$ denotes the differential operator obtained by Taylor expanding $f(X)$ and replacing each term by the corresponding differential operator. Note that $j$ is an even function and that $j(0)=1$, so that the operator \eqref{duffing} is a sum of operators that starts with the one obtained via symmetric quantization. It follows that
\begin{align}
    \mathcal{D}(e^{-i\langle \cdot, X\rangle}) 
    =
    j^{1/2}(X)e^{-i\langle \hat p, X\rangle}
    \label{dufloexp}
\end{align}
for the special case of plane waves \eqref{expo}.

\paragraph{Plane waves and lone star lemma.} One can define the momentum representations as in sections \ref{ssec: momentum rep 1}--\ref{ssec: momentum rep 2}, now using Duflo ordering instead of the symmetric one above. In particular, the star product is defined as
\begin{align}
    A\star B 
    :=
    \mathcal{D}^{-1}(\hat A \hat B),
\end{align}
with an ordering that differs from that of eq.~\eqref{eq: def star product}. Exponential operators now act as
\begin{align}
    e^{-i\langle \hat p, X\rangle}\Phi(p) = \frac{e^{-i\langle p, X\rangle}}{j^{1/2}( X)}\star \Phi(p)
    \label{dufloexpop}
\end{align}
owing to eq.~\eqref{dufloexp}. They still get composed thanks to Baker-Campbell-Hausdorff formula \eqref{eq: BCH}, so one has
\begin{align}
    \frac{e^{-i\langle p, X\rangle}}{j^{1/2}( X)}\star\frac{e^{-i\langle p, Y\rangle}}{j^{1/2}( Y)}= \frac{e^{-i\langle p, B(X,Y)\rangle}}{j^{1/2}( B(X,Y))}.
    \label{duflostarbch}
\end{align}
Thanks to \eqref{dufloexpop}, the projector \eqref{proji} onto the $\mathcal{I}$-invariant subspace reads
\begin{align}
    \hat P_\mathcal I\cdot\Phi(p) 
    =
    \frac{1}{|\mathcal I|}\sum_{X\in \operatorname{Logs(e)}}\frac{e^{-i\langle p, X\rangle}}{j^{1/2}( X)}\star \Phi(p).
\end{align}
The position representations of section \ref{sec: representations} are unaffected. It is thus straightforward to adapt the derivation \eqref{eq: ode g of exp}--\eqref{e63} to find that noncommutative plane waves now read
\begin{align}
    E(X,p)=\frac{e^{-i\langle p, X\rangle}}{j^{1/2}(X)}.
    \label{duflopw}
\end{align}
Their $\mathcal{I}$-invariant counterparts \eqref{s16} are
\begin{align}
    E_\mathcal{I}(X,p)
    =
    \frac{1}{|\operatorname{Logs}(\exp(X))|}\sum_{Y\in \operatorname{Logs(\exp(X))}}\frac{e^{-i\langle p, Y\rangle}}{j^{1/2}(Y)}.
    \label{dufloipw}
\end{align}
The definitions of Fourier transforms and series still read as in \eqref{eq: def ncft} and \eqref{ourdef}, and their adjoints are also defined as in \eqref{defad}. The difference is that plane waves are given by \eqref{duflopw} and \eqref{dufloipw} instead of \eqref{e85}--\eqref{e86}. Almost all the properties of section \ref{sec:DefFourier} hold without change, as they are nearly all based on the algebraic fact that star products of plane waves represent the Baker-Campbell-Hausdorff group law \eqref{duflostarbch}. The only properties that get modified are the lone star lemma \eqref{eq: lone star lemma} and the explicit expressions \eqref{e85}--\eqref{e86} of $\mathcal{I}$-invariant plane waves. Let us start by adapting the former.

\noindent\i{Statement.} With the Duflo quantization map \eqref{duffing}, the lone star lemma reads
\begin{align}
      \int_{\mathfrak{g}^*} \frac{\dd^n p}{(2\pi)^n}\;\bar \Phi\star \Psi(p) 
     =
     \int_{\mathfrak{g}^*} \frac{\dd^n p}{(2\pi)^n}\; \Delta^{1/2}\left(-i\frac{\partial}{\partial p}\right)\overline{\Phi(p)} \Psi(p),
     \label{eq: duflo lone star lemma}
\end{align}
where $\Delta^{1/2}(X) = e^{\tr \mathrm{ad}_{X/2}}$ is the square root of the modular function. In particular, if $G$ is unimodular, the lone star lemma holds {\it strictly}, with no differential operator inside the integral. As a direct consequence, the inverse Fourier transform \eqref{defad} can be recast as
\begin{align}
      \cF_{\mathcal{I}}^{\dagger}[\Phi](g)
      =
      \sqrt{|\mathbb Z|^r}\frac{e^{\tr \mathrm{ad}_{X/2}}}{j^{1/2}(\log(g))}\int_{\mathfrak{g}^*}\frac{\dd^np}{(2\pi)^n} e^{i\langle p, \log(g)\rangle}\Phi(p).
      \label{eq: duflo no star inv Fourier}
\end{align}
without any star product, nor any differential operator.
\begin{center}
\begin{minipage}{.9\textwidth}
{\footnotesize%
\i{Proof.} Since the star product is bilinear, and since the Fourier transform is bijective, it suffices to show the property for noncommutative plane waves, which provide a basis of functions in momentum space. This yields
\begin{align*}
      \int_{\mathfrak{g}^*} \frac{\dd^n p}{(2\pi)^n}\;\overline{E(X,p)}\star E(Y,p) 
= 
\int \frac{\dd^n p}{(2\pi)^n}\;E(B(-X,Y),p) 
=
\delta_\mathfrak{g}(B(-X,Y)).
\end{align*}
Now using the fact that the Haar delta function for the Baker-Campbell-Hausdorff group is related to the Lebesgue delta function by a Jacobian, find
\begin{equation*}
\delta_\mathfrak{g}(B(-X,Y))
=
\frac{\delta^d(X-Y)}{J(X)}
=
\frac{\delta^d(X-Y)e^{\tr \mathrm{ad}_{X/2}}}{j^{1/2}(X)j^{1/2}(Y)}
=
\int_{\mathfrak{g}^*} \frac{\dd^n p}{(2\pi)^n}\; \Delta^{1/2}\left(-i\frac{\partial}{\partial p}\right)\!\overline{E(X,p)} E(Y,p),
\end{equation*}
where we also used the fact that $j$ is an even function of $X$.\hfill$\blacksquare$}
\end{minipage}
\end{center}

Regarding the explicit expression of $\mathcal{I}$-invariant plane waves, let us adopt the same notations as around \eqref{e84}. Let us also make an assumption similar to the one below \eqref{e116}, namely that $j^{1/2}(X)$ can be expressed as a ratio $j^{1/2}(X) = \frac{n(X)}{d(X)}$, with a function $n(X)$ that is $2\pi a_i(X)$-periodic and $d(X)$ a polynomial or a square root of a polynomial. These would respectively be $\sin\|X\|$ and $\|X\|$ for SU(2). Then, from eq.~\eqref{dufloipw}, one has
\begin{align}
    E_\mathcal{I}(X,p)
    &=
    \frac{1}{|\mathbb Z|^r n(X)}\sum_{\mathbf{n}\in \mathbb Z^r} d(X+2\pi n^i a_i(X)) e^{-i\langle p, X+2\pi n^i a_i(X)\rangle}\\
    &=
    \frac{d(i\partial_p)}{|\mathbb Z|^r n(X)}\sum_{\mathbf{n}\in \mathbb Z^r}e^{-i\langle p, X+2\pi n^i a_i(X)\rangle}
    \\
    &=
    \frac{d(i\partial_p)}{n(X)}\left(e^{-i\langle p,X\rangle}\sum_{\mathbf{k}\in \mathbb Z^r}\prod_{i=1}^r\frac{\delta(\langle p, a_i(X)\rangle - k_i)}{|\mathbb Z|}\right),
    \label{e614}
\end{align}
which holds for almost any $X\in\mathfrak{g}$. This mildly differs from the $\cal I$-invariant plane wave \eqref{e85} that stems from symmetric ordering. In fact, the Fourier coefficients for the Duflo and Gutt star products are related by
\begin{equation}
    F_\mathcal{I}^{\text{Duflo}}[\psi](p) 
    =
    j^{-1/2}\!\left(i\frac{\partial}{\partial p}\right) F_\mathcal I^{\text{Gutt}}[\psi](p).
    \label{eRelGuttDuflo}
\end{equation}

\paragraph{SU(2) characters.} As an explicit comparison between Duflo-ordered and symmetric-ordered Fourier transforms, consider the example of SU(2) characters (recall section \ref{ssec: SU(2)}). Using \eqref{eBANGER ABSOLU} and the relation \eqref{eRelGuttDuflo} between the two Fourier transforms, pseudodifferential operators simplify so that, in the end,
\begin{align}
    F_\mathcal{I}^{\text{Duflo}}[\chi_\lambda](p) 
    =
    \frac{2\pi^2}{\sqrt{|\mathbb Z|}} \frac{\delta(\|p\|-(2\lambda+1))}{\|p\|}.
\end{align}
This striking result suggest that the Duflo-ordered Fourier transform of an irreducible character is localized and constant on the coadjoint orbit that corresponds to Kirillov's character formula. It is, in fact, possible to prove this starting from the formula \eqref{eq: duflo no star inv Fourier} for the inverse Fourier transform, applied to unimodular groups. Indeed, on the one hand, the character is given by the inverse Fourier transform
\begin{equation}
    \chi_\lambda(g)
      =
      \sqrt{|\mathbb Z|^r}\frac{1}{J^{1/2}(\log(g))}\int_{\mathfrak{g}^*}\frac{\dd^np}{(2\pi)^n} e^{i\langle p, \log(g)\rangle}F_\mathcal{I}[\chi_\lambda](p).
      \label{ea17}
\end{equation}
On the other hand, Kirillov's character formula states that
\begin{align}
    \chi_\lambda(g) 
    =
    \frac{1}{J^{1/2}(\log(g))}\int \frac{\dd^n p}{(2\pi)^n} e^{i\langle p, \log(g)\rangle}\delta_{\mathcal{O}_{\lambda + \rho}}(p),
    \label{ea18}
\end{align}
where $\rho$ denotes the half-sum of positive roots, and $\delta_{\mathcal{O}_{\lambda + \rho}}$ is the delta distribution on the orbit. More precisely, for any test function $\Psi(p)$, one has
\begin{align}
\int \frac{\dd^np}{(2\pi)^n}\Psi(p)\delta_{\mathcal{O}_{\lambda + \rho}}(p)
=
\int_{\mathcal{O}_{\lambda + \rho}} \Psi\big\vert_{\mathcal{O}_{\lambda + \rho}} \mu_{\lambda + \rho},
\end{align}
with $\Psi\vert_{\mathcal{O}_{\lambda + \rho}}$ the restriction of $\Psi$ to the orbit. The Liouville volume form $\mu_{\lambda + \rho}$ of the orbit $\mathcal{O}_{\lambda + \rho}$ is assumed to be normalized to $\dim \lambda$, the dimension of the representation $\lambda$.  Comparing the right-hand sides of \eqref{ea17} and \eqref{ea18}, one identifies
\begin{align}
    F_\mathcal{I}[\chi_\lambda](p) 
    =
    \frac{1}{\sqrt{|\mathbb Z|^r}}\delta_{\mathcal O_{\lambda+\rho}}(p).
\end{align}
The Duflo-ordered Fourier transform thus relates characters of irreducible representations to coadjoint orbits. In this sense, the Duflo-ordered noncommutative Fourier transform establishes a manifest correspondence between functions on the group and functions localized on a discrete set of coadjoint orbits. Note that using the newly proposed $\mathcal{I}$-invariant plane waves \eqref{dufloipw} was crucial to derive such a correspondence. This is because, in the SU(2) case, the delta localization of characters can be traced back to the delta distributions appearing in \eqref{e614}. 

Finally, one can also use the Duflo-ordered Fourier transform to derive the Parseval-Plancherel identity. Indeed, let $\psi\in L^2(G)$ be a class function, meaning that $\psi(h g h^{-1}) = \psi(g)$ for all $h$ and $g$ in $G$. Any class function on $G$ can be written as a sum of characters, so $\psi = \sum_{\lambda\in \hat G} c_\lambda[\psi] \chi_\lambda(g)$, where $c_\lambda[\psi]$ is computed thanks to the orthonormality of characters: 
\begin{align}
    \langle \chi_\lambda|\psi\rangle_G 
    =
    \int \bar\chi_\lambda(g)\psi(g) 
    =
    c_\lambda[\psi].
\end{align}
Now, take any two class functions $\phi$ and $\psi$. Since the noncommutative Fourier transform is an isometry, one has $\langle \phi,\psi\rangle_G = \langle F_\mathcal I[\phi],F_\mathcal I[\psi]\rangle_{\mathfrak g^*}$. Explicitly writing this last scalar product in $L^2_\star(\mathfrak{g}^*)$ and using the lone star lemma \eqref{eq: duflo lone star lemma}, one finds
\begin{align}
    \langle F_\mathcal{I}[\phi]|F_\mathcal{I}[\psi]\rangle_{\mathfrak{g}^*} 
    &= 
    \sum_{\lambda,\sigma \in \hat G} 
    \bar c_\lambda[\phi]c_\sigma[\psi] 
    \int_{\mathfrak{g}^*} \frac{\dd^n p}{(2\pi)^n}
    \sqrt{\delta_{\mathcal{O_{\lambda + \rho}}}(p)\delta_{\mathcal{O}_{\sigma + \rho}}(p)} 
    \\
    &=
    \sum_{\lambda\in \hat G} \bar c_\lambda[\phi]c_\lambda[\psi]\int_{\mathcal{O_{\lambda+\rho}}}\mu_{\lambda + \rho} 
    =
    \sum_{\lambda\in \hat G} \dim \lambda\;\bar c_\lambda[\phi]c_\lambda[\psi],
\end{align}
which leads to the Parseval-Plancherel identity, $\langle \phi|\psi\rangle_G =\sum_{\lambda\in \hat G} \dim \lambda\;\bar c_\lambda[\phi]c_\lambda[\psi]$. Recall, by contrast, that no such simplification was available around eqs.~\eqref{e107bis}--\eqref{planche} for noncommutative Fourier series with symmetric ordering instead of Duflo ordering.

\section*{Acknowledgements}

We are grateful to Sylvain Carrozza for first bringing noncommutative Fourier transforms to our attention. We also thank Isma\"el Ahlouche Lahlali, Glenn Barnich, Guillaume Bossard, Pierre Delplace, Benoit Estienne, Romane Houvenaghel and Alberto Nardin for motivating discussions on related topics.

\addcontentsline{toc}{section}{References}

\providecommand{\href}[2]{#2}
\end{document}